\documentclass[dvipsnames,twocolumn]{aastex631}
\usepackage{graphicx}   
\usepackage{lineno}
\usepackage{bm}

\shorttitle{Cosmology with one galaxy?}
\shortauthors{Villaescusa-Navarro et al.}
\graphicspath{{./}{figures/}}

\begin{document}

\title{Cosmology with one galaxy?}

\author[0000-0002-4816-0455]{Francisco Villaescusa-Navarro}
\affiliation{Center for Computational Astrophysics, Flatiron Institute, 162 5th Avenue, New York, NY, 10010, USA}
\affiliation{Department of Astrophysical Sciences, Princeton University, Peyton Hall, Princeton NJ 08544, USA}

\author{Jupiter Ding}
\affiliation{Department of Astrophysical Sciences, Princeton University, Peyton Hall, Princeton NJ 08544, USA}

\author{Shy Genel}
\affiliation{Center for Computational Astrophysics, Flatiron Institute, 162 5th Avenue, New York, NY, 10010, USA}
\affiliation{Columbia Astrophysics Laboratory, Columbia University, New York, NY, 10027, USA}

\author{Stephanie Tonnesen}
\affiliation{Center for Computational Astrophysics, Flatiron Institute, 162 5th Avenue, New York, NY, 10010, USA}

\author{Valentina La Torre}
\affiliation{Department of Physics and Astronomy, Tufts University, Medford, MA 02155, USA}

\author{David N. Spergel}
\affiliation{Center for Computational Astrophysics, Flatiron Institute, 162 5th Avenue, New York, NY, 10010, USA}
\affiliation{Department of Astrophysical Sciences, Princeton University, Peyton Hall, Princeton NJ 08544, USA}

\author{Romain Teyssier}
\affiliation{Department of Astrophysical Sciences, Princeton University, Peyton Hall, Princeton NJ 08544, USA}

\author[0000-0002-0701-1410]{Yin Li}
\affiliation{Center for Computational Astrophysics, Flatiron Institute, 162 5th Avenue, New York, NY, 10010, USA}
\affiliation{Center for Computational Mathematics, Flatiron Institute, 162 5th Avenue, New York, NY, 10010, USA}

\author{Caroline Heneka}
\affiliation{University of Hamburg, Hamburg Observatory, Gojenbergsweg 112, 21029 Hamburg, Germany}

\author[0000-0002-4728-8473]{Pablo Lemos}
\affiliation{Department of Physics  and Astronomy, University of Sussex,Brighton, BN1 9QH, UK}
\affiliation{University College London, Gower St, London, UK }

\author[0000-0001-5769-4945]{Daniel Angl\'es-Alc\'azar}
\affiliation{Department of Physics, University of Connecticut, 196 Auditorium Road, Storrs, CT, 06269, USA}
\affiliation{Center for Computational Astrophysics, Flatiron Institute, 162 5th Avenue, New York, NY, 10010, USA}

\author{Daisuke Nagai}
\affiliation{Department of Physics, Yale University, New Haven, CT 06520, USA}

\author{Mark Vogelsberger}
\affiliation{Kavli Institute for Astrophysics and Space Research, Department of Physics, MIT, Cambridge, MA 02139, USA}

\correspondingauthor{Francisco Villaescusa-Navarro}
\email{fvillaescusa@flatironinstitute.org}

\begin{abstract}
Galaxies can be characterized by many internal properties such as stellar mass, gas metallicity, and star-formation rate. We quantify the amount of cosmological and astrophysical information that the internal properties of individual galaxies and their host dark matter halos contain. We train neural networks using hundreds of thousands of galaxies from 2,000 state-of-the-art hydrodynamic simulations with different cosmologies and astrophysical models of the CAMELS project to perform likelihood-free inference on the value of the cosmological and astrophysical parameters. We find that knowing the internal properties of a single galaxy allow our models to infer the value of $\Omega_{\rm m}$, at fixed $\Omega_{\rm b}$, with a $\sim10\%$ precision, while no constraint can be placed on $\sigma_8$. Our results hold for any type of galaxy, central or satellite, massive or dwarf, at all considered redshifts, $z\leq3$, and they incorporate uncertainties in astrophysics as modeled in CAMELS. However, our models are not robust to changes in subgrid physics due to the large intrinsic differences the two considered models imprint on galaxy properties. We find that the stellar mass, stellar metallicity, and maximum circular velocity are among the most important galaxy properties to determine the value of $\Omega_{\rm m}$. We believe that our results can be explained taking into account that changes in the value of $\Omega_{\rm m}$, or potentially $\Omega_{\rm b}/\Omega_{\rm m}$, affect the dark matter content of galaxies. That effect leaves a distinct signature in galaxy properties to the one induced by galactic processes. Our results suggest that the low-dimensional manifold hosting galaxy properties provides a tight direct link between cosmology and astrophysics.
\end{abstract}


\section{Introduction} 
\label{sec:intro}

The discovery that the Universe is accelerating its expansion \citep{perlmutter_supernovae, supernovae1998} marked an inflexion point in cosmology. Determining the nature and properties of the substance responsible for this behaviour, dark energy, is one of the most important goals of current cosmology.

In order to accomplish this task we need to extract the maximum information from cosmological surveys. We know that for Gaussian density fields the power spectrum (or the correlation function) is the optimal estimator to extract the maximum available information. However, we do not know what estimator would allow us to extract the maximum amount of information for non-Gaussian density fields, like the matter and galaxy distribution resembles on non-linear scales. Quantifying the information content from different estimators is currently a very active area of research \citep{Quijote, Samushia_2021, Gualdi_2021, Kuruvilla_2021, Bayer_2021,  Banerjee_2019, Changhoon_2019, Uhlemann_2020, Friedrich_2020, Massara_2020, Dai_2020, Allys_2020, Banerjee_2020, Banerjee_2021, Gualdi_2020, Gualdi_2021, Giri_2020, Bella_2020, Changhoon_2020, Valgiannis_2021, Bayer_2021, Kuruvilla_2021b, Naidoo_2021, Porth_2021, Harnois_2021, Liu_2019, Zack_2019, Will_2019, Marques_2019, Lee_2020a, Lee_2020, Gemma_2019, Ajani_2020, Harnois_2021a, Sihao_2021, Harnois_2021}.

Another avenue is to use machine learning techniques, e.g. neural networks, to find an approximation to the optimal estimator \citep{Siamak_16, Schmelzle_17, Gupta_18, Ribli_19, Fluri_19, Ntampaka_19, Sultan_2019, Jose_2020, Niall_2020, Paco_2021b, Lu_2021}. Recent works have shown that even for fields that are very contaminated by astrophysical effects, it is possible to extract cosmological information from small scales \citep{Paco_2021a}.

Either way, both approaches yield to the same conclusion: small, non-linear, scales seem to contain a wealth of cosmological information. Down to which scale is this statement true? Does the information run out at some point? These are difficult questions that we do not attempt to address in this work. Instead, we investigate whether there is any cosmological information in one of the fundamental blocks of many cosmological surveys: galaxies. In other words, can we infer the value of the cosmological parameters from a single, generic, galaxy?

To address this question we employ machine learning methods to connect the internal properties of individual galaxies to the value of the cosmological and astrophysical parameters. We made use of galaxies from the state-of-the-art hydrodynamic simulations of the CAMELS project \cite{CAMELS}. The internal galaxy properties considered in this work include the stellar mass, the star-formation rate, the total mass in the galaxy's subhalo, and the stellar radius among others. The CAMELS simulations contain around 1 million galaxies at fixed redshift for 2,000 different cosmological and astrophysical models from two completely different suites of hydrodynamic simulations. This allows us to quantify the dependence of our results on uncertainties in astrophysics processes and on the subgrid physics model.

As we shall see below, we find that we can infer the value of $\Omega_{\rm m}$ with a $\simeq10\%$ precision just using the internal properties of an individual, generic, galaxy; no constraint can be placed on the value of $\sigma_8$. These constraints account for uncertainties in astrophysics, as implemented on the CAMELS simulations. However, they are not robust to changes in subgrid physics, due to the intrinsic differences in galaxy properties between the different simulations. If our interpretation of these results is correct, it would imply that galaxy properties live in manifolds that change with the value of $\Omega_{\rm m}$, providing a link between cosmology and astrophysics. To our knowledge, this is the first time that this idea has been explored and it may open new exciting possibilities to connect cosmology with astrophysics through galaxy properties.

To enable the community to reproduce our results we release all data used in this work together with the codes, databases, and network weights obtained after training. We refer the reader to \url{https://github.com/franciscovillaescusa/Cosmo1gal} for further details.

This paper is organized as follows. In Sec. \ref{sec:methods} we describe the data we use together with the machine learning methods employed. We present our results in Sec. \ref{sec:results} and attempt a physical interpretation for them in Sec. \ref{sec:interpretation}. Finally, we summarize and discuss the main results of this work in Sec. \ref{sec:summary}.

\section{Methods} 
\label{sec:methods}

In this section we describe the data and the machine learning models we use to find the mapping between the properties of individual galaxies and the value of the cosmological and astrophysical parameters.

\subsection{Simulations}

In this work we use galaxies from the simulations of the CAMELS project \citep{CAMELS}. CAMELS contains two different suites of state-of-the-art hydrodynamic simulations: 
\begin{itemize}
\item \textbf{IllustrisTNG.} The simulations in this suite have been run with the \textsc{AREPO} code \citep{Arepo_public} and employ the same subgrid physics model as the original IllustrisTNG simulations \citep{Pillepich_2018, IllustrisTNG_public}.
\item \textbf{SIMBA.} The simulations in this suite have been run with the \textsc{GIZMO} code \citep{Hopkins2015_Gizmo} and employ the same subgrid physics model as the original SIMBA simulation \citep{SIMBA}, building on its precursor MUFASA \citep{Dave2016} with the addition of supermassive black hole growth and feedback \citep{Angles-Alcazar2017_BHfeedback}.
\end{itemize}
All simulations follow the evolution of $2\times256^3$ dark matter plus fluid elements in a periodic comoving volume of $(25~h^{-1}{\rm Mpc})^3$ from $z=127$ down to $z=0$. All simulations share the value of these cosmological parameters: $\Omega_{\rm b}=0.049$, $h=0.6711$, $n_s=0.9624$, $\sum m_\nu=0.0$ eV, $w=-1$. However, each simulation has a different value of $\Omega_{\rm m}$ and $\sigma_8$. The hydrodynamic simulations also vary the values of four astrophysical parameters that control the efficiency of supernova and active galactic nuclei (AGN) feedback: $A_{\rm SN1}$, $A_{\rm SN2}$, $A_{\rm AGN1}$, and $A_{\rm AGN2}$.

In this work we use the LH sets of the IllustrisTNG and SIMBA suites. Each set contains 1,000 simulations, where the value of $\Omega_{\rm m}$, $\sigma_8$, $A_{\rm SN1}$, $A_{\rm SN2}$, $A_{\rm AGN1}$, and $A_{\rm AGN2}$ are arranged in a latin-hypercube defined by
\begin{eqnarray}
\Omega_{\rm m}&\in&[0.1, 0.5]\\
\sigma_8&\in&[0.6, 1.0]\\
A_{\rm SN1}, A_{\rm AGN1}&\in&[0.25, 4.0]\\
A_{\rm SN2}, A_{\rm AGN2}&\in&[0.5, 2.0]~,
\end{eqnarray}
and each simulation has a different value of the initial random seed. We note that the latin-hypercubes of the IllustrisTNG and SIMBA simulations are different, i.e. there is no correspondence between simulations among the two sets. We emphasize that the astrophysics parameters have very different meanings in the IllustrisTNG vs SIMBA suites.

We refer the reader to \cite{CAMELS} for further details on the simulations of the CAMELS project.

\subsection{Galaxy properties}

We have run \textsc{SUBFIND} \citep{subfind} to identify halos and subhalos in the simulations. In this work we consider galaxies as subhalos that contain more than 20 star particles. All galaxies from all simulations are characterized by 14 different properties:
\begin{enumerate}
\item \bm{$M_{\rm g}$}. The gas mass content of the galaxy, including the contribution from the circumgalactic medium.
\item \bm{$M_{\rm BH}$}. The black-hole mass of the galaxy.
\item \bm{$M_*$}. The stellar mass of the galaxy.
\item \bm{$M_{\rm t}$}. The total mass of the subhalo hosting the galaxy, i.e. the sum of the mass in dark matter, gas, stars, and black-holes in the subhalo.
\item \bm{$V_{\rm max}$}. The maximum circular velocity of the subhalo hosting the galaxy:  $V_{\rm max}=\max(\sqrt{GM(<R)/R}$).
\item \bm{$\sigma_v$}. The velocity dispersion of all particles contained in the galaxy's subhalo.
\item \bm{$Z_{\rm g}$}. The mass-weighted gas metallicity of the galaxy.
\item \bm{$Z_*$}. The mass-weighted stellar metallicity of the galaxy.
\item \bm{${\rm SFR}$}. The galaxy star-formation rate.
\item \bm{$J$}. The modulus of the galaxy's subhalo spin vector.
\item \bm{$V$}. The modulus of the galaxy's subhalo peculiar velocity.
\item \bm{$R_*$}. The radius containing half of the galaxy stellar mass.
\item \bm{$R_{\rm t}$}. The radius containing half of the total mass of the galaxy's subhalo.
\item \bm{$R_{\rm max}$}. The radius at which $\sqrt{GM(<R_{\rm max})/R_{\rm max}}=V_{\rm max}$.
\end{enumerate}
For galaxies of the IllustrisTNG simulations we also consider three additional properties:
\begin{enumerate}
 \setcounter{enumi}{14}
\item \bm{${\rm U}$}. The galaxy magnitude in the U band.
\item \bm{${\rm K}$}. The galaxy magnitude in the K band.
\item \bm{${\rm g}$}. The galaxy magnitude in the g band.
\end{enumerate}
We note that the reason why the latter properties are only present in the IllustrisTNG simulations is because the version of \textsc{SUBFIND} we employed does not account for the differences in the simulation suites and therefore it cannot estimate these magnitudes.

We have employed the above galaxy properties because these are computed by \textsc{SUBFIND} and therefore easily accessible to us. Using other properties would require post-processing the snapshots; we leave this for future work. We emphasize that while most of the considered properties can be associated to galaxies themselves, there are others that should be seen as properties of the subhalos hosting the galaxies, like $V_{\rm max}$, $M_t$, and $\sigma_v$. We note that in this work we are not splitting galaxies according to some property, e.g. large or small, central or satellite.

\subsection{Machine learning algorithms}

We made use of several machine learning algorithms for three different reasons. First, to assert that our conclusions hold independently of the method used. Second, because some tasks (e.g. feature ranking) require a significant amount of computation and is difficult to perform them if not using fast methods. And third, to provide some interpretability to our results.

\begin{itemize}

\item \textbf{Gradient Boosting Trees}: This method is based on decision trees and therefore computationally efficient. We made use of the XGB package\footnote{\url{https://xgboost.readthedocs.io}} to estimate the value of $\Omega_{\rm m}$ from the 17 galaxy properties. For each task we tune the value of the following hyperparameters: 1) the learning rate, 2) the maximum depth, 3) the minimum child weight, 4) the value of gamma, 5) the colsample\_bytree, and 6) the number of estimators. The loss function we optimize is the mean squared error. XGB accounts for L2 regularization internally. We note that in this case we perform parameter regression, while with neural networks we do likelihood-free inference.

\item \textbf{Neural networks}: We made use of fully connected layers since they are appropriate for the task we consider in this work. Our architecture consists of several fully connected layer blocks. These blocks contain a fully connected layer that is followed by a LeakyReLU activation layer with a slope of 0.2, and a dropout layer where the value of the dropout rate is a hyperparameter. The very last layer of the architecture is just a fully connected layer not followed by an activation or dropout layer. The hyperparameters we consider are: 1) the number of fully connected layers, 2) the number of neurons in each layer, 3) the dropout value, 4) the value of the weight decay, and 5) the value of the learning rate. Our networks are trained to perform likelihood-free inference; they estimate the posterior mean and standard deviation for each parameter by minimizing the loss function of moment networks \citep{moment_networks}.

\end{itemize}

\begin{figure*}[t]
\centering
\includegraphics[width=0.99\linewidth]{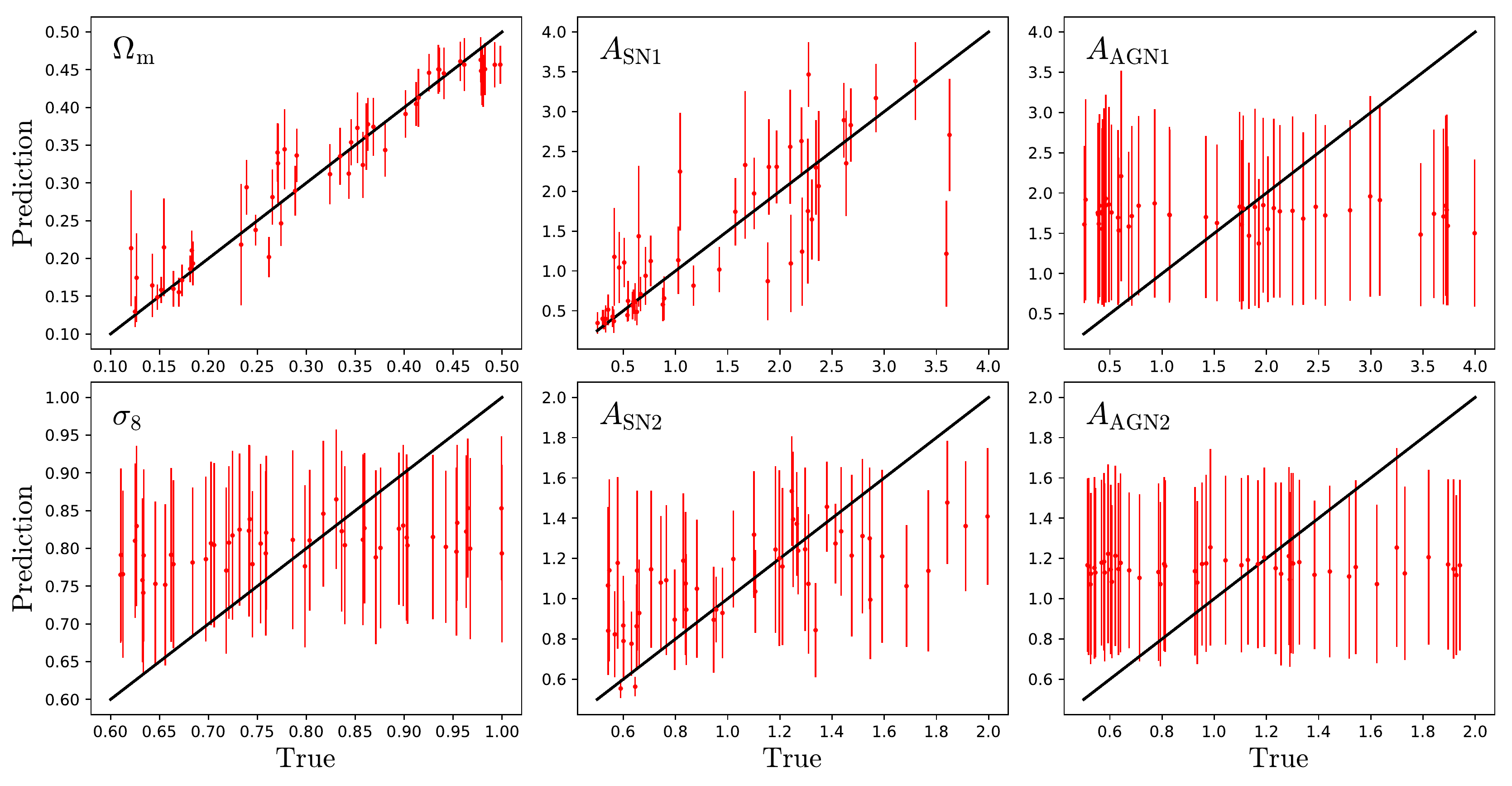}
\caption{We have trained a neural network to perform likelihood-free inference on the value of the cosmological ($\Omega_{\rm m}$ and $\sigma_8$) and astrophysical ($A_{\rm SN1}$, $A_{\rm SN2}$, $A_{\rm AGN1}$, and $A_{\rm AGN2}$) parameters using as input 17 properties of individual galaxies from the IllustrisTNG simulations at $z=0$. Once the network is trained, we test it using individual galaxies from the test set. The different panels show the posterior mean and standard deviation predicted by the network versus the true value. Every point with its errorbar represents a single galaxy. We find that our model is able to infer the value of $\Omega_{\rm m}$ from the properties of individual galaxies with a $\sim10\%$ precision.}
\label{fig:results_IllustrisTNG}
\end{figure*} 

We use the optuna\footnote{\url{https://optuna.org}} \citep{Optuna} package to perform the hyperparameter optimization of both gradient boosting trees and neural networks. In both cases, we first sample the hyperparameter space using between 25 to 30 trials\footnote{A trial represent the result of training the model with a given value of the hyperparameters.} and then we perform Bayessian optimization for 75-80 more trials. In all cases we search the hyperparameter space to minimize the value of the validation loss.

For gradient boosting trees and neural networks we split the data into three different sets: training, validation, and testing. Since galaxies in the same simulations may share features in either low or high-dimensional spaces, we first split the data by simulation. The training set contains 850 simulations with all their galaxies. The validation set has all galaxies from 100 simulations, while the testing set contains 50 simulations with all their galaxies. By splitting the data in this way, we can guarantee that the galaxies in the test set, together with their associated cosmology and astrophysics, have never been seen by the model before.

\subsection{Accuracy and precision}

Throughout this paper we will be discussing the accuracy and the precision of a given model. Here we describe what we mean by these.

\textbf{Neural networks.} With this method we perform likelihood-free inference, and the output of the networks is the posterior mean ($\mu$) and standard deviation ($\sigma$) of a given parameter $i$, i.e.
\begin{eqnarray}
\mu_i(\textbf{X}) &=& \int_{\theta_i} p_i(\theta_i | \textbf{X}) \theta_i d\theta_i~,\\
&& \nonumber \\
\sigma_i(\textbf{X}) &=& \int_{\theta_i} p_i(\theta_i | \textbf{X}) (\theta_i - \mu_i)^2 d\theta_i~,
\label{Eq:mean_posterior}
\end{eqnarray}
where $\textbf{X}$ is the vector containing the galaxy properties and $p_i(\theta_i|\textbf{X})$ is the marginal posterior over the parameter $i$
\begin{equation}
p_i(\theta_i|\textbf{X}) = \int_{\theta} p_i(\theta_i | \textbf{X})d\theta_1...d\theta_{i-1}d\theta_{i+1}...d\theta_n~.
\label{Eq:variance_posterior}
\end{equation}
We define the accuracy of the model for the parameter $i$ as
\begin{equation}
{\rm Accuracy}_i=\sqrt{\left\langle (\theta_i - \mu_i)^2 \right\rangle}~, 
\end{equation}
where $\theta_i$ is the true value of the parameter $i$ and the average is done over all galaxies in the considered set (e.g. the test set). Meanwhile, we define the precision of the model on the parameter $i$ as
\begin{equation}
{\rm Precision}_i=\left\langle \frac{\sigma_i}{\mu_i} \right\rangle    
\end{equation}
\textbf{Gradient boosting trees}. With this method we only perform parameter regression, and the output of the model is the predicted value of the parameter $i$: $\tilde{\theta}_i$. In this case we only define the model accuracy:
\begin{equation}
    {\rm Accuracy}_i=\sqrt{\left\langle (\theta_i - \tilde{\theta}_i)^2 \right\rangle}~.
\end{equation}
We emphasize the differences between our definitions of accuracy and precision. Accuracy quantifies the dispersion around the true values (independently of the size of the error bars for the prediction), while precision estimates the size of the relative errors (independently on whether the values are close or far from the true values).

Finally, we note that the accuracy and precision as defined above will give more weight to low-mass galaxies, as those are the most abundant in the simulations. When studying how constraints change for different galaxies in a given simulation we can quantify the dependence of accuracy and precision on stellar mass.

\section{Results}
\label{sec:results}

We start by training a neural network that takes as input the 17 properties of individual galaxies of the IllustrisTNG simulations at $z=0$ and outputs the posterior mean and standard deviation for each cosmological and astrophysical parameter. Once the network is trained, we test it using the properties of individual galaxies of the test set. In Fig. \ref{fig:results_IllustrisTNG} we show the derived posterior means and standard deviations for 50 random galaxies versus their true value.

\begin{figure*}[t]
\centering
\includegraphics[width=0.99\linewidth]{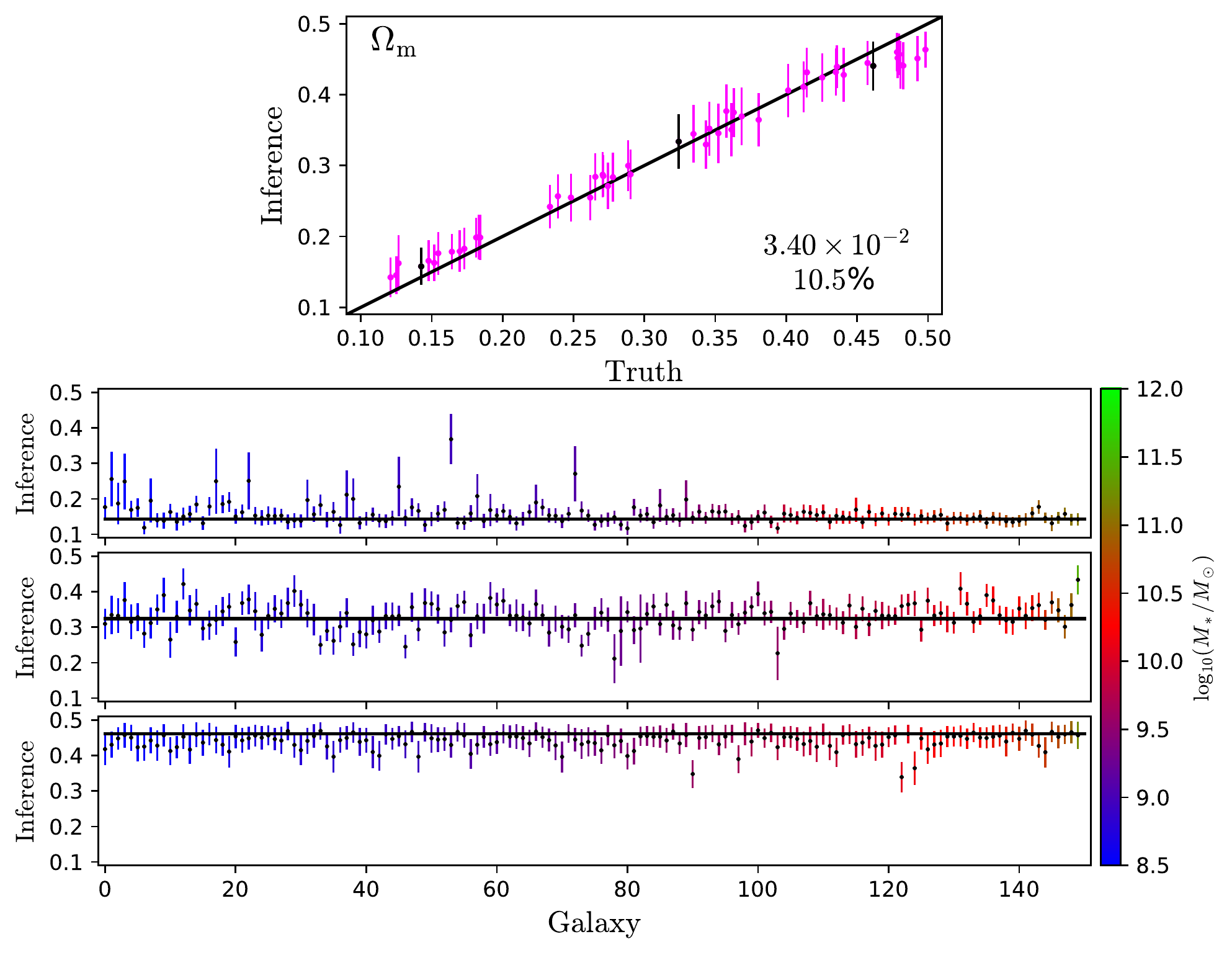}
\caption{We trained neural networks using galaxies from 850 IllustrisTNG simulations, and have reserved all the galaxies from 50 additional IllustrisTNG simulations for the test set. For each galaxy of a given simulation of the test set we compute the posterior mean and standard deviation. The bottom panels show the results for 150 individual galaxies of three different simulations with three different values of $\Omega_{\rm m}$ (shown with a horizontal solid line) color coded according to the value of the stellar mass of the galaxy. Galaxies are organized according to their stellar mass; galaxies on the left are small while the ones on the right are large. We have then computed the posterior mean and standard deviation from all galaxies in a simulation (Eq. \ref{Eq:mean_values}) and plot the results in the top panel. The black points in that panel show the results for the simulations in the bottom panels. The numbers inside the top panel show the accuracy and precision of the model. All results are at $z=0$. As can be seen, our network is able to infer the value of $\Omega_{\rm m}$ for the vast majority of galaxies in a given simulation.}
\label{fig:IllustrisTNG_big}
\end{figure*} 

The network has not found enough information to infer the value of $A_{\rm AGN1}$, $A_{\rm AGN2}$, and $\sigma_8$, so it just predicts the mean value with large errorbars for these parameters. For the supernova parameters, $A_{\rm SN1}$ and $A_{\rm SN2}$, the network may be using some information to provide some loose constraints (we provide further details in the appendix \ref{sec:astro_params}). On the other hand, for $\Omega_{\rm m}$, the network seems to have found enough information to determine its value for almost all galaxies considered. We emphasize that these constraints are derived for individual galaxies, each having a different cosmology and astrophysics model, that were selected randomly, i.e. independently of their stellar mass and whether they are centrals or satellites.

From Fig. \ref{fig:results_IllustrisTNG} we cannot tell whether the network is able to infer the value of $\Omega_{\rm m}$ for any generic galaxy or whether we were lucky with the random selection we carried out in that exercise. To shed light on this question we compute the average mean and standard deviation of the posterior for all galaxies in a simulation of the test set,
\begin{equation}
\bar{\mu}_i = \frac{1}{N}\sum_{j=1}^N \mu_{i,j} \hspace{1cm} \bar{\sigma} = \frac{1}{N}\sum_{j=1}^N \sigma_{i,j}~,
\label{Eq:mean_values}
\end{equation}
where $i$ denotes the considered parameter (e.g. $\Omega_{\rm m}$) and $j$ runs over all $N$ galaxies of a given simulation. In the top panel of Fig. \ref{fig:IllustrisTNG_big} we show the above values for each of the simulations in the test set. In the bottom right part of that panel we quote the accuracy and precision of the model. As can be seen, on average for all galaxies, the network is able to infer the value of $\Omega_{\rm m}$ with an accuracy of $0.034$ and a  $10.5\%$ precision.

\begin{figure*}[t]
\centering
\includegraphics[width=0.99\linewidth]{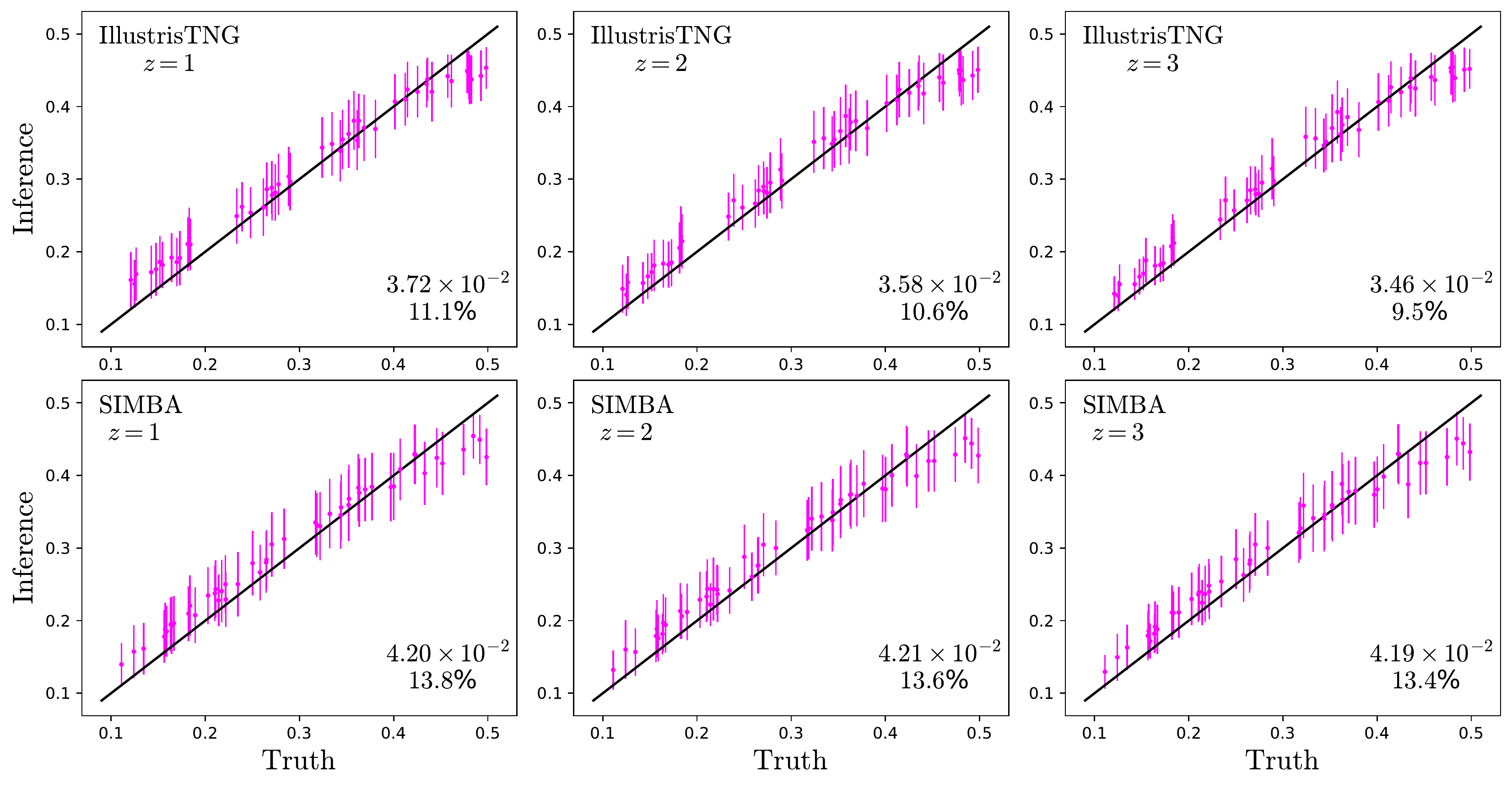}
\caption{Redshift dependence. We have trained neural networks to infer the value of the cosmological and astrophysical parameters using properties of individual galaxies at different redshifts and for galaxies of the IllustrisTNG and SIMBA simulations. For each galaxy of each simulation of the test set we compute the posterior mean and standard deviation for $\Omega_{\rm m}$. Next, we compute the mean of those two numbers (Eq. \ref{Eq:mean_values}) and plot them in the figure for the 50 different simulations in the test set. We show results at redshifts 1, 2, and 3. The numbers in the bottom right corner show the model accuracy and precision. As can be seen, our networks can infer the value of $\Omega_{\rm m}$ from individual galaxies at redshifts higher than $z=0$ with an accuracy similar to the one achieved by the models at $z=0$.}
\label{fig:redshift_evolution}
\end{figure*}

We perform the following exercise to investigate in more detail whether our model works for all galaxies or just a subset of them. First, we select three different simulations of the test set with different values of $\Omega_{\rm m}$: one low, one high, and one intermediate. From each of those simulations we randomly select 150 galaxies. For each of those galaxies we compute the posterior mean and standard deviation of $\Omega_{\rm m}$. In the bottom panels of Fig. \ref{fig:IllustrisTNG_big} we show the results. The constraints are color-coded according to the stellar mass of the galaxies. Those plots show that our network not only works for a subset of galaxies, but seems to perform well for the majority of the galaxies in a given simulation.

Three features are worth noticing. First, in all cases there seems to be some outliers where the posterior mean is significantly away from the true value. Second, for the models with intermediate and high values of $\Omega_{\rm m}$, the size of the standard deviation of the posterior is very similar for all galaxies\footnote{For the model with high $\Omega_{\rm m}$, the minimum and maximum values only vary by a factor of $\sim3$, while for the model with low $\Omega_{\rm m}$ the difference is more than a factor of $\sim7$.}, while for the cosmology with a low value of $\Omega_{\rm m}$ we find that massive galaxies have smaller posterior variances than low-mass galaxies. Third, from the top panel of Fig. \ref{fig:IllustrisTNG_big} we can see that in some simulations there seems to be systematic differences between the posterior means and the true value. We will attempt to provide an explanation for these features in Sec. \ref{sec:interpretation}.

From the above results we conclude that there is evidence showing that the value of $\Omega_{\rm m}$ can be inferred from the properties of individual galaxies for the vast majority of the cases. This statement holds for galaxies with very different cosmologies, astrophysics, and almost independently on whether the galaxy is massive or dwarf, central or satellite\footnote{We note that we never provided the models with information on whether the galaxies are centrals or satellites}...etc. 

\begin{figure*}[t]
\centering
\includegraphics[width=0.80\linewidth]{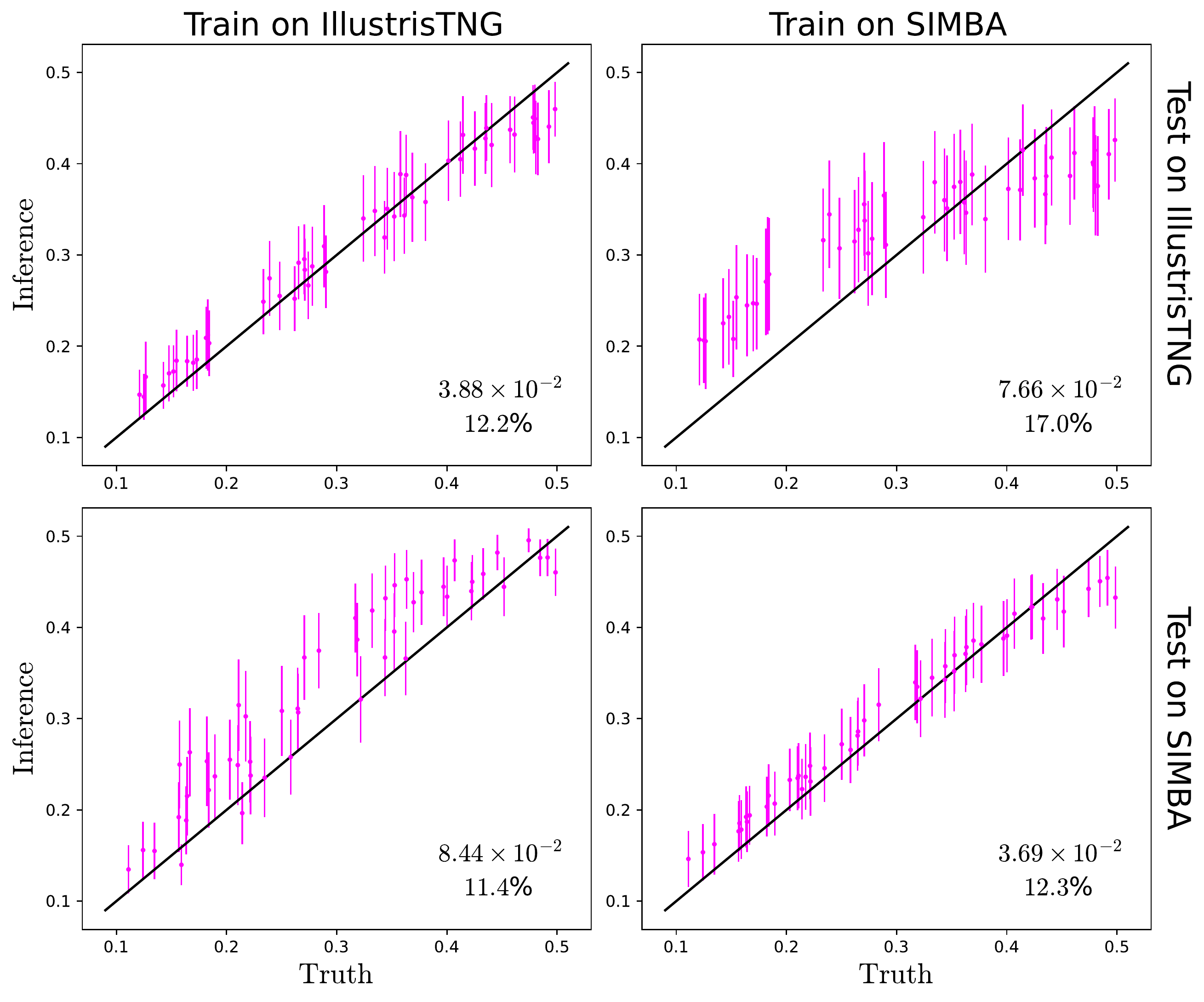}
\caption{Robustness test. We have trained neural networks to perform likelihood-free inference on the value of the cosmological and astrophysical parameters using internal properties of individual galaxies at $z=0$. In this case we made use of the 14 internal properties that are common between the galaxies in the IllustrisTNG and SIMBA simulations. We have trained models using galaxies from either the IllustrisTNG or SIMBA simulations. For each simulation in the test set, we compute the posterior mean and standard deviation for $\Omega_{\rm m}$ for each galaxy on it. We then compute the average value of those two numbers from all galaxies in a given simulation (Eq. \ref{Eq:mean_values}). These panels show the results for all 50 simulations in the test set when training on galaxies of a given simulation and test it on galaxies of the same simulation or another simulation. In the bottom right part of each panel we quote the accuracy and precision of the model on the tested galaxies. As can be seen, when the model is tested on galaxies from simulations different to the ones used for training, the model is not able to infer the correct cosmology in most of the cases. This indicates that the model is not robust and may be using information that is specific to each galaxy formation model.}
\label{fig:robustness}
\end{figure*}

\subsection{Dependence on method and data}

We have carried out other sanity checks to investigate whether our conclusions hold for different methods and different simulations:
\begin{itemize}
\item We have repeated the above analysis but using galaxies from the CAMELS-SIMBA simulations (using their 14 properties) instead of the IllustrisTNG galaxies, reaching the same conclusions as above. We provide further details of this test and its results in the appendix \ref{sec:SIMBA}. 
\item We have repeated the above exercise but performing parameter regression through gradient boosting trees. We have trained these models using both galaxies from the IllustrisTNG and SIMBA simulations. We find that the accuracy of these methods on $\Omega_{\rm m}$ is similar to the one from neural networks. 
\end{itemize}
These tests indicate that our results are robust to the particularities of the method used to perform the mapping between galaxy properties and the value of $\Omega_{\rm m}$.

\subsection{Dependence on redshift}

We now investigate whether our results only hold at $z=0$ or we can also infer the value of $\Omega_{\rm m}$ from internal properties of galaxies at higher redshifts. For this, we have trained neural networks to infer the value of the cosmological and astrophysical parameters from galaxies at redshifts 1, 2, and 3 using both the IllustrisTNG and the SIMBA simulations.

Once the models are trained, we test it on individual galaxies from simulations of the test set, and compute the average posterior mean and posterior standard deviation from all galaxies in a given simulation (i.e. Eq. \ref{Eq:mean_values}). We then show these measurements in Fig. \ref{fig:redshift_evolution}.

\begin{figure*}[t]
\centering
\includegraphics[width=0.49\linewidth]{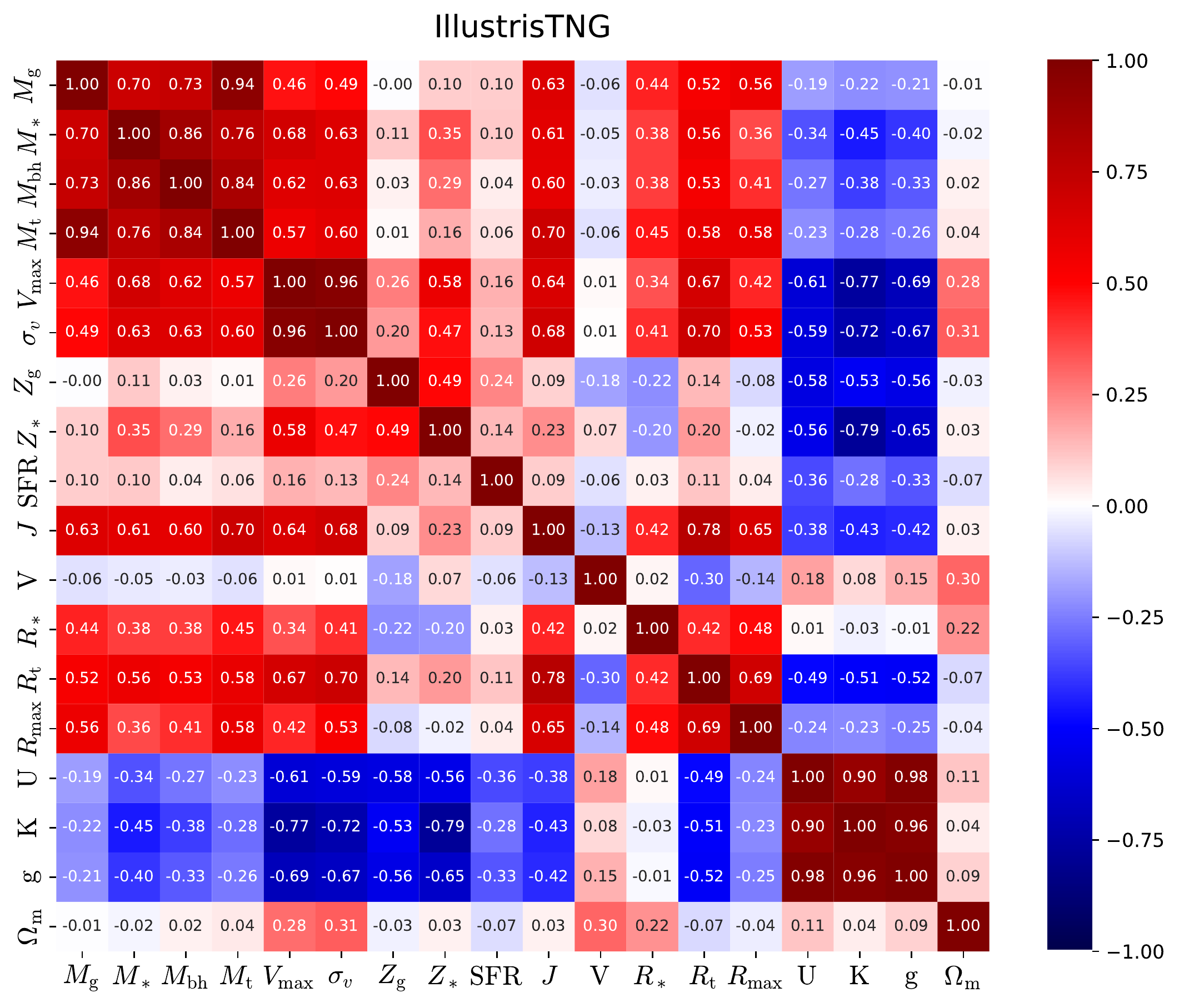}
\includegraphics[width=0.49\linewidth]{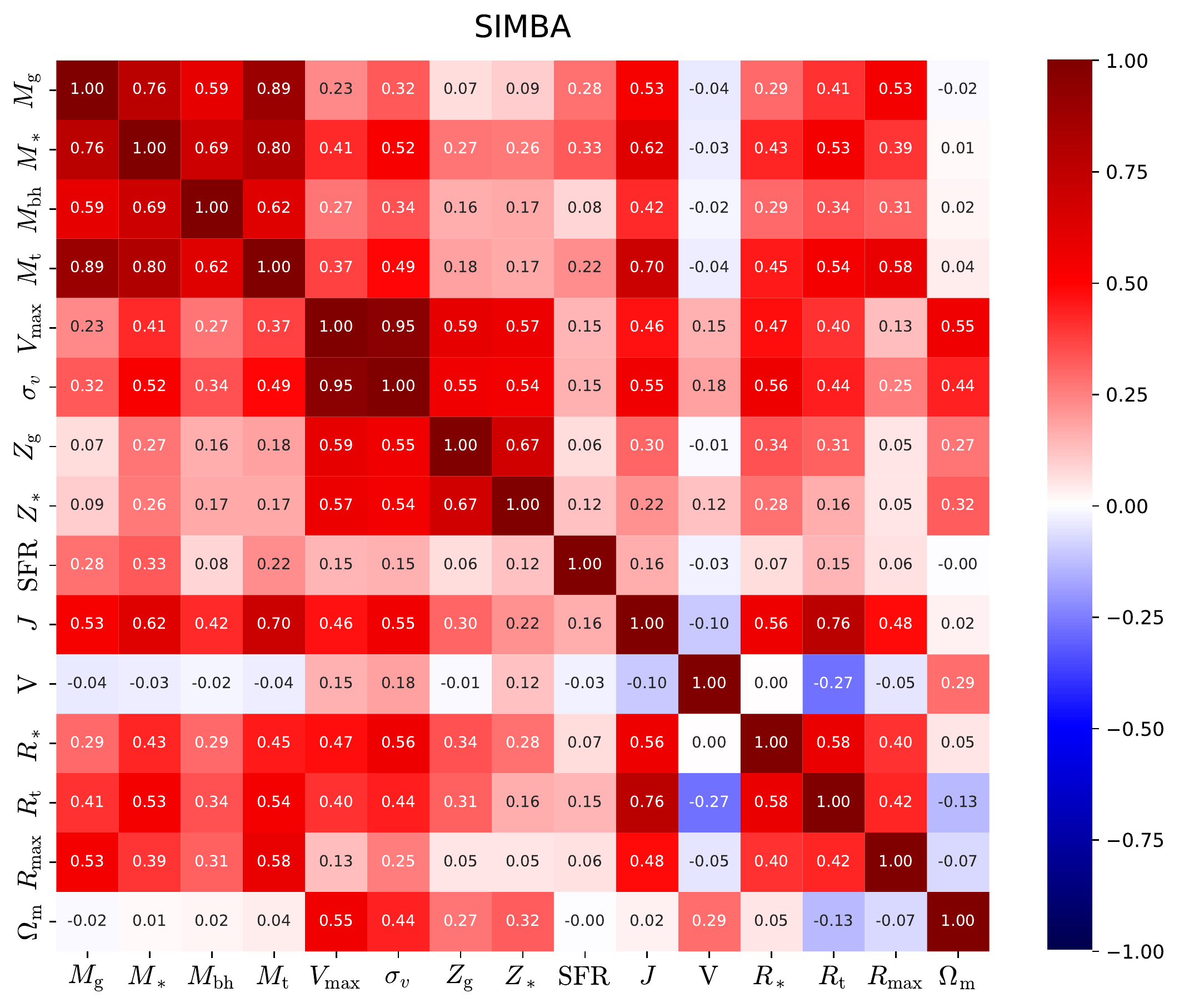}
\caption{We have computed the correlation matrix (see Eq. \ref{Eq:pearson}) of the galaxy properties plus $\Omega_{\rm m}$ for the IllustrisTNG (left) and SIMBA (right) simulations. We find strong linear correlation among different galaxy properties (e.g. gas mass and total mass), but the correlations between $\Omega_{\rm m}$ and the galaxy properties are relatively mild. This indicates that the value of $\Omega_{\rm m}$ cannot be inferred due to simple, linear correlations between $\Omega_{\rm m}$ and galaxy properties.}
\label{fig:correlation_matrix}
\end{figure*} 

As can be seen, results at redshifts higher than zero are qualitatively very similar to the ones at $z=0$, for both IllustrisTNG and SIMBA galaxies. For all models we have computed their accuracy and precision, and we quote them in the bottom right part of each panel. We find that both the accuracy and precision of the models is very similar across redshifts, although there is a slight improvement when using galaxies at higher redshfits. The models trained on IllustrisTNG galaxies exhibit however a better accuracy and precision than the ones trained on SIMBA galaxies. This is due to the inclusion of the three additional features contained in the IllustrisTNG galaxies (the magnitudes in the U, K, and g bands). This can be seen more clearly in Fig. \ref{fig:robustness}, where models trained on the same variables from IllustrisTNG and SIMBA exhibit a similar accuracy and precision. Overall, we conclude that it seems possible to infer the value of $\Omega_{\rm m}$ from internal properties of galaxies at redshifts $z\leq 3$.

Next, we investigate whether our results are independent of redshift, i.e. whether a model trained on galaxies at a given redshift is able to infer the value of $\Omega_{\rm m}$ from galaxies at a different redshift. We have tried this on different models at different redshifts and found that it does not work. We have also tried a few different things to verify that the reason was not due to the use of comoving versus proper quantities \citep[see e.g.][]{Helen_2021} but we did not find any improvement. From these tests we conclude that the mapping between the internal galaxy properties and $\Omega_{\rm m}$ should have an intrinsic dependence on redshift. In the next section we attempt to provide a physical understanding of this result.

\subsection{Robustness}
\label{subsec:robust}

Ideally, we would like to apply this method to internal properties of real galaxies to derive the value of $\Omega_{\rm m}$ and see whether it agrees with the one derived from standard cosmological measurements (e.g. CMB or galaxy clustering). However, to carry out that task, we need a robust model, i.e. that it works independently of the type of simulations used for training. At its core, CAMELS was designed to test the robustness of models by providing simulations from two completely different suites: IllustrisTNG and SIMBA. 

Here we quantify the robustness of our model by testing the models trained on IllustrisTNG and SIMBA galaxies on galaxies from the SIMBA and IllustrisTNG simulations, respectively. We show the results of such exercise in Fig. \ref{fig:robustness}. We find that while testing the model on galaxies from the same subgrid model as the one used for training yields precise and accurate results for both the IllustrisTNG and SIMBA models, the model fails when tested on galaxies from different subgrid models. In the appendix \ref{sec:appendix_robustness} we provide further details on this test.

We have repeated this exercise with the gradient boosting tree method reaching the same conclusions. We have also tried with a smaller set of variables, e.g. $\{M_*, V_{\rm max}, Z_* \}$, but the models are still not robust. We thus conclude that our models may be learning something particular about each simulation or that the two different simulations do not overlap in parameter space. In the next section we shall see that one reason behind this behaviour is that the two different suites of simulations produce very different galaxies with distinct properties, limiting the range where they both overlap and therefore making the model not robust.

\section{Interpretation}
\label{sec:interpretation}

In this section we attempt to provide a physical explanation to our findings above. We will focus our attention on  $\Omega_{\rm m}$.

\subsection{Linear correlations}
\label{sec:lincorr}

We start by investigating whether there are strong linear correlations between the galaxy properties and $\Omega_{\rm m}$. For that, we plot in Fig. \ref{fig:correlation_matrix} the correlation matrix of all galaxy properties plus $\Omega_{\rm m}$, defined as
\begin{equation}
    R_{ij}=\frac{C_{ij}}{\sqrt{C_{ii}C_{jj}}}
\label{Eq:pearson}
\end{equation}
where
\begin{equation}
    C_{ij}=\langle (p_i -\bar{p}_i)(p_j-\bar{p}_j)\rangle
\end{equation}
where $p_i$ refers to the $i$ feature of the data vector (galaxy properties plus $\Omega_{\rm m}$) and $\bar{p}_i=\langle p_i \rangle$. This matrix gives information about the linear correlation between the different variables. 

As can be seen, while some galaxy properties seem to be highly correlated (e.g. $V_{\rm max}$ and $\sigma_v$) the linear correlations between $\Omega_{\rm m}$ and the galaxy features are not particularly high. For IllustrisTNG galaxies, the strongest correlated variable with $\Omega_{\rm m}$ is $\sigma_v$, while for SIMBA galaxies is $V_{\rm max}$. 

These tests indicate that our findings are not due to simple linear correlations between $\Omega_{\rm m}$ and galaxy properties.

As a side calculation we have also carried out an analysis with the Principal Component Analysis (PCA) to try to identify the number of components and variables that are responsible for most of the overall data variance (i.e. considering both galaxy properties plus $\Omega_{\rm m}$). For IllustrisTNG galaxies we find that the first principal component is dominated by $\Omega_{\rm m}$ as well as $V$, $V_{\rm max}$ and $\sigma_v$, while for SIMBA the most important features are $\Omega_{\rm m}$ and $V_{\rm max}$, followed by $Z_{\rm g}$, $Z_*$, $\sigma_v$, and $V$. It is interesting to see that $\Omega_{\rm m}$ and $V_{\rm max}$ seem to form a basic to explain most of the data variance.

\subsection{Properties ranking}
\label{subsec:features}

Next, we try to identify the most important galaxy features that the network is using in order to carry out the inference. We have used different methods to perform this task, like computing saliency maps and SHAP values for the neural networks, and using the \textit{feature importance} method for random forest and gradient boosting trees regressors. However, we found that these methods did not allow us to identify the most important features; likely because of the strong internal correlations between the different variables. In the Appendix \ref{sec:SHAP} we provide additional details about our results when using SHAP values.

\begin{figure*}[t]
\centering
\includegraphics[width=0.49\linewidth]{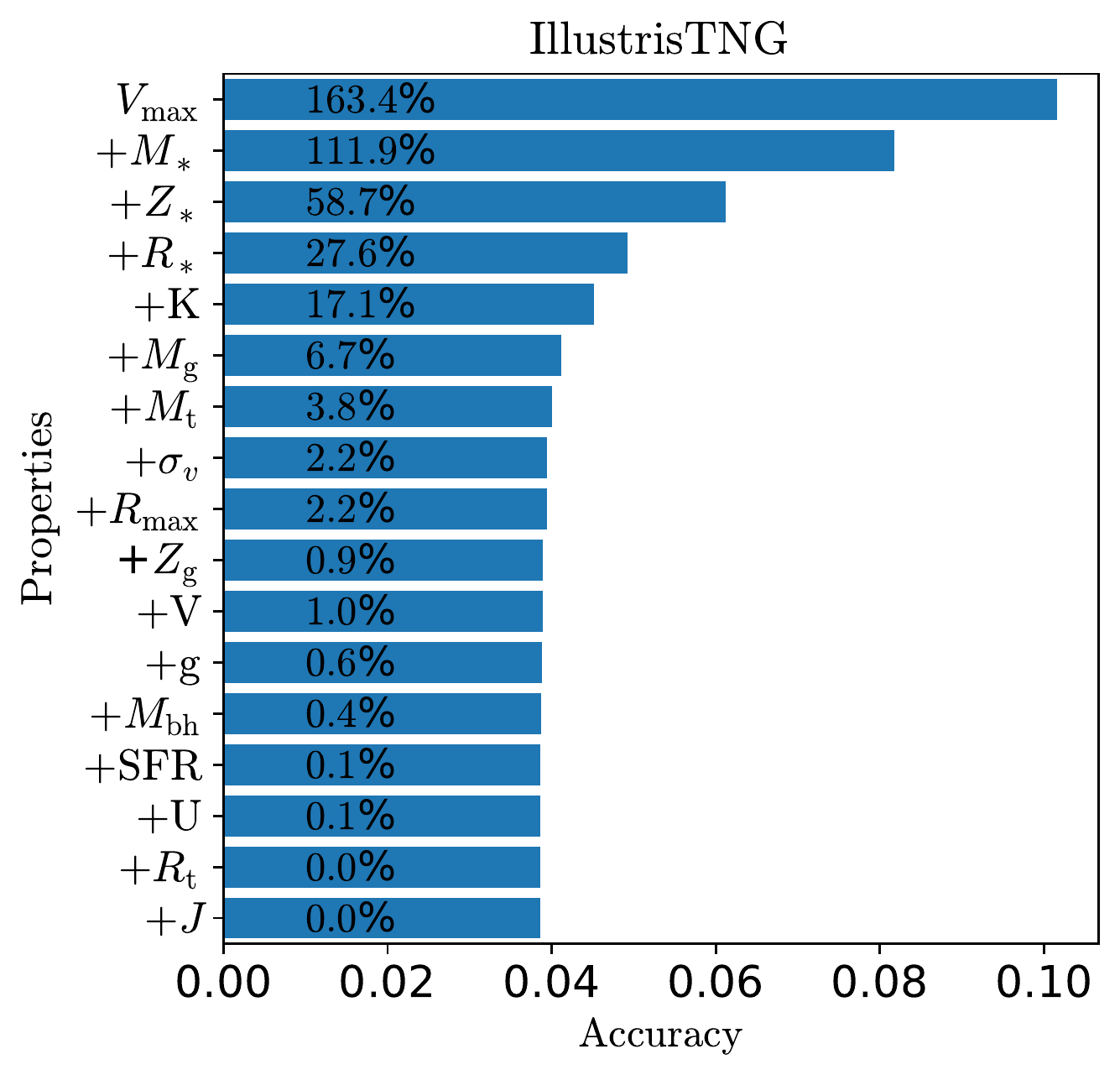}
\includegraphics[width=0.49\linewidth]{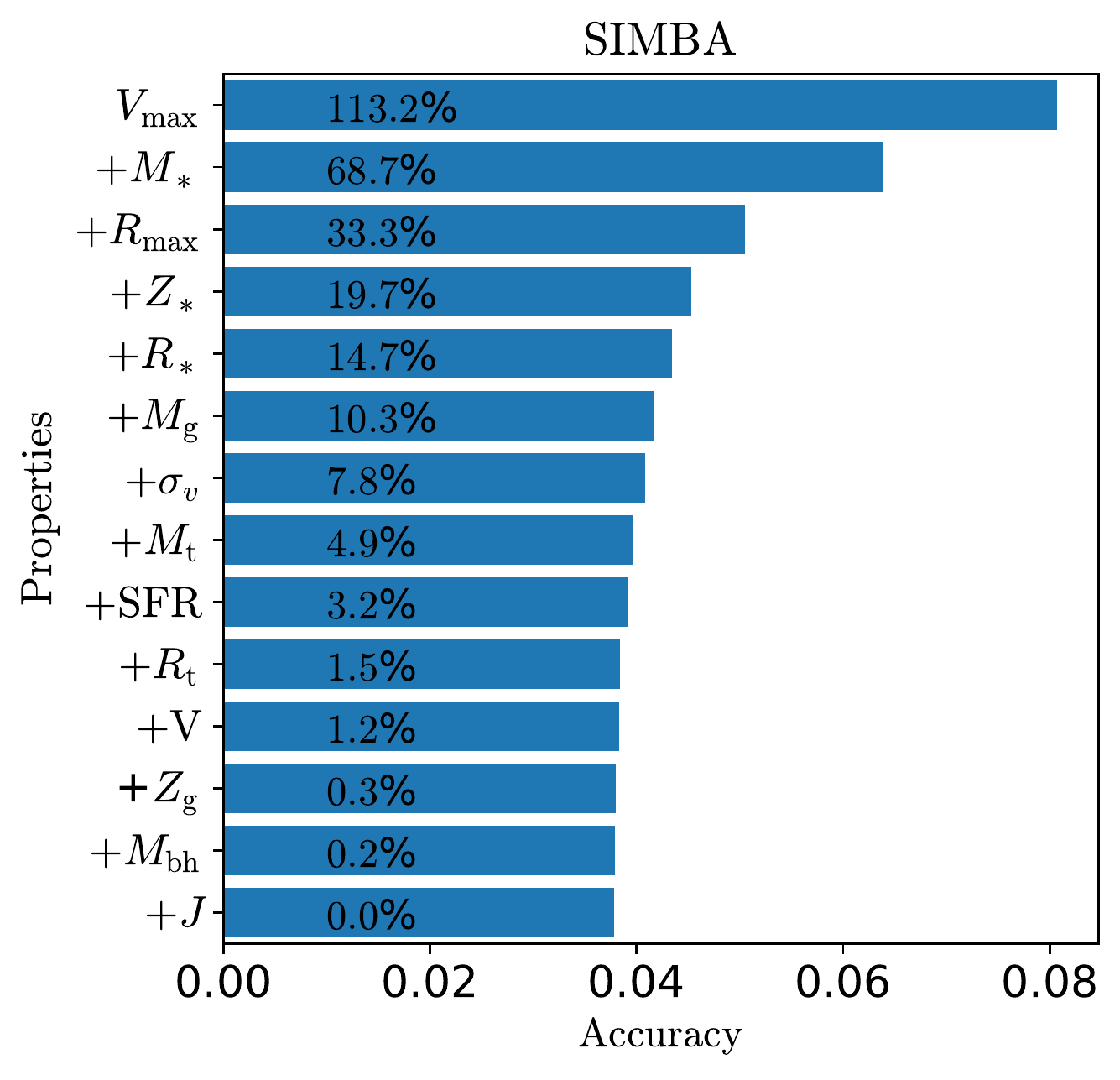}
\caption{We rank order the galaxy properties for both IllustrisTNG (left) and SIMBA (right) such that the variables contributing the most to the model accuracy are on top while the features contributing the least are on the bottom (see text for details on the procedure used). The horizontal bars indicate the accuracy (in terms of RMSE) achieved by the considered variables and the black numbers inside them show the loss in accuracy with respect to a model trained using all variables. For instance, for the IllustrisTNG galaxies, a model that only uses $V_{\rm max}$ achieves a RMSE of $\sim0.1$ and performs 163.4\% worse than the model trained on all 17 properties. Likewise, a model trained on SIMBA galaxies using $\{V_{\rm max}, M_*, R_{\rm max}, Z_*, R_*\}$ achieves a RMSE of $\simeq0.04$, which is only 14.7\% worse than the model trained on all 14 galaxy properties. We emphasize that this ordering was derived when training gradient boosting trees models to perform regression to the value of $\Omega_{\rm m}$.}
\label{fig:ranking}
\end{figure*}

We tackle this problem as follows. First, we train a model using all galaxy properties and record its accuracy. Next, we remove one of the considered properties and retrain a model using the rest of properties. We then reincorporate that feature, remove another property, and train another the model on those variables. We repeat this procedure until all properties have been removed. For instance, we train a model that contains all properties except gas mass, we train another model that contains all properties except stellar mass, we train another model that contains all properties except black-hole mass and so on. For each model we save the accuracy obtained. This method allows us to quantify the worsening of the model accuracy by removing a single feature.

We then continue the exercise by removing the variable that changes the accuracy the least. With the subset of variables left, we repeat the above procedure and train models where we remove one galaxy property at a time and record the model accuracy. In this way we can rank order\footnote{We note that this method is not guaranteed to give the correct ordering in general. For instance, removing two or more properties at a time may lead to a different ordering.} the features according to their contribution to the model accuracy. Unfortunately, doing this exercise with neural networks while performing hyperparameter optimization is too computationally expensive for this work, so we decided to do it using gradient boosting trees instead of neural networks.

We show the rank ordered features in Fig. \ref{fig:ranking}. We find the two most important features to be $V_{\rm max}$ and $M_*$ for both IllustrisTNG and SIMBA galaxies. The stellar metallicity and stellar radius are also among the five most important features in both cases. However, for IllustrisTNG galaxies, the K-band seems a very relevant property (this property is not present in the SIMBA galaxies) while in the case of SIMBA galaxies the radius associated to the maximum circular velocity, $R_{\rm max}$, is selected as an important feature. In Fig. \ref{fig:ranking} we show the accuracy (quantified in terms of root mean squared error) gained as we add variables. For IllustrisTNG galaxies, using $\{V_{\rm max}, M_*, Z_*, R_*, {\rm K} \}$ only degrade results by 17\% with respect to the accuracy achieved by training on all 17 properties. Meanwhile, for SIMBA galaxies, using $\{V_{\rm max}, M_*, R_{\rm max}, Z_*, R_* \}$ only degrades results by 15\% with respect to training using all 14 features. 

Next, with these subsets of variables we have trained neural networks to perform likelihood-free inference. For IllustrisTNG/SIMBA galaxies we find that the accuracy on predicting $\Omega_{\rm m}$ degrades by $27\%$/28\% when comparing it to the accuracy of model trained using all 17/14 variables. When using the 5 most important features according to the absolute SHAP values (e.g. $\{ M_*, {\rm K}, M_{\rm g}, Z_{\rm g}, V_{\rm max} \}$ for the IllustrisTNG simulations) we found that the model performs significantly worse: the root mean squared error between the posterior mean and the true value degrades by 47\%. These results show that this procedure can find a minimum set of variables that is responsible for most of the model accuracy. 

\begin{figure*}[t]
\centering
\includegraphics[width=0.99\linewidth]{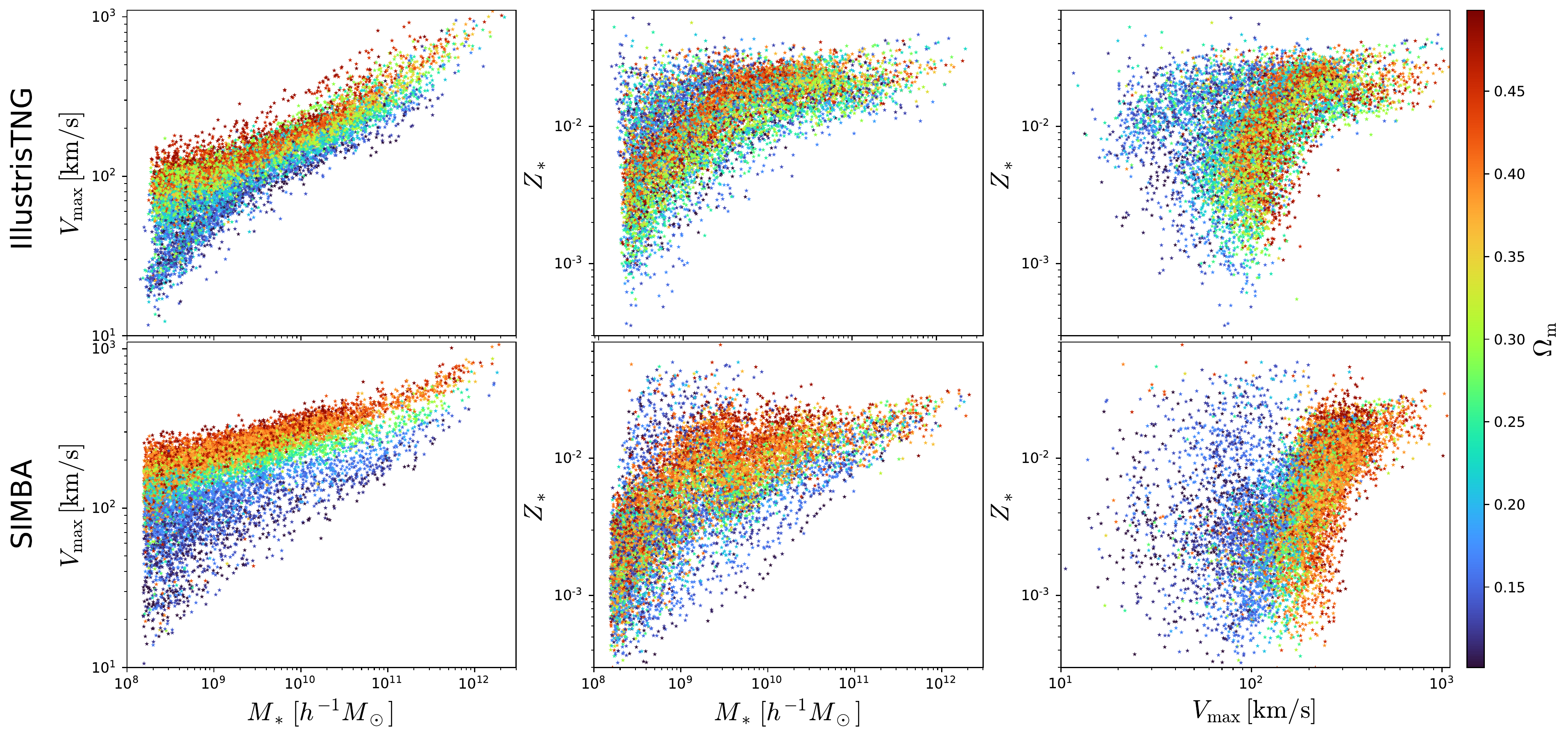}
\includegraphics[width=0.99\linewidth]{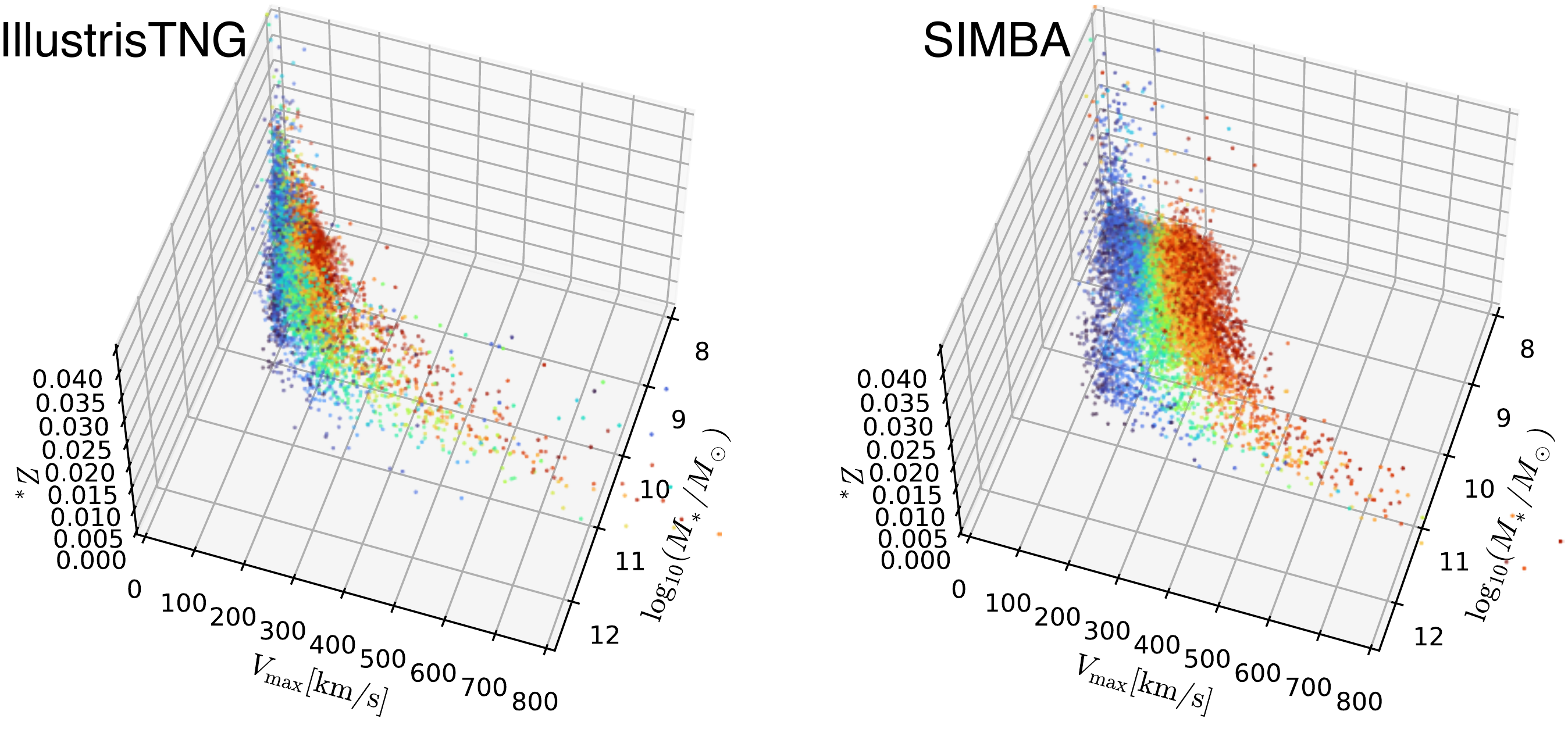}
\caption{For both the IllustrisTNG and SIMBA suites we  have randomly taken 100 simulations. From each simulation we have randomly selected 100 galaxies at $z=0$, for a total of 10,000 galaxies. \textbf{Top.} For each of those galaxies we show correlations between $V_{\rm max}$ and $M_*$ (left), $Z_*$ and $M_*$ (middle), and $Z_*$ and $V_{\rm max}$ (right) for the IllustrisTNG (top row) and SIMBA (bottom row) galaxies. Each galaxy is color coded according to the value of $\Omega_{\rm m}$ of its simulation (blue/green/red indicate low/medium/high values of $\Omega_{\rm m}$). As can be seen, there is a prominent correlation between $V_{\rm max}$ and $M_*$ that changes with $\Omega_{\rm m}$. We can also observe other more complex trends in the $Z_*-M_*$ and $Z_*-V_{\rm max}$ planes with $\Omega_{\rm m}$. \textbf{Bottom:} We show the results in 3D. Galaxies occupy different regions in the properties space depending on their value of $\Omega_{\rm m}$. We believe that in higher dimensions (i.e. considering more galaxy properties) galaxies should occupy even more disconnected regions as a function of $\Omega_{\rm m}$. We interpret these results as  $\Omega_{\rm m}$ changing the manifold where galaxy properties reside in a different way as feedback does. Machine learning methods can use these patterns to determine the value of  $\Omega_{\rm m}$.}
\label{fig:Mstar_Vmax_Zstar}
\end{figure*}

\subsection{Visual inspection}

Before attempting a physical explanation of our results with the information gained from the above experiments, we perform a visual inspection of some galaxy features in 2 and 3 dimensions to gain intuition. For this, we randomly select 10,000 galaxies from 100 different IllustrisTNG simulations (100 galaxies per simulation); we do the same exercise for the SIMBA simulations. For this exercise we consider three galaxy properties: $V_{\rm max}$, $M_*$, and $Z_*$. We have chosen these variables because $V_{\rm max}$ and $M_*$ are the most important ones for both IllustrisTNG and SIMBA galaxies, while $Z_*$ is the among the four\footnote{We note that $Z_*$ is the third and fourth most important variable for IlllustrisTNG and SIMBA galaxies, respectively. The third more important variable for SIMBA is $R_{\rm max}$, that is not among the most important variables for IllustrisTNG.} more important variables in both suites. 

In Fig. \ref{fig:Mstar_Vmax_Zstar} we show 2D and 3D projections of the data. Each point, representing a galaxy, is color coded according to its $\Omega_{\rm m}$ value. As can be seen, galaxy properties occupy different regions in the 2D and 3D plots depending on the value of $\Omega_{\rm m}$. In particular, the dependence of the $V_{\rm max} - M_*$ relation on $\Omega_{\rm m}$ is particularly pronounced. We will discuss this trend in more detail in the next subsection. We emphasize that galaxies are randomly selected from the simulations, i.e. they not only differ on the value of $\Omega_{\rm m}$ but also on $\sigma_8$ and on the values of the four astrophysical parameters considered. 

From Fig. \ref{fig:Mstar_Vmax_Zstar} we can also see the large, intrinsic differences between the SIMBA and IllustrisTNG galaxies: while they exhibit similar qualitative dependence with $\Omega_{\rm m}$, they populate the parameter space differently. This is however expected, given the large differences between the IllustrisTNG and SIMBA subgrid models. We note that in higher dimensions, the differences between the simulations may be even more pronounced. We believe that this is the reason why our models are not robust; i.e. a model trained on galaxy properties from IllustrisTNG simulations does not work when tested on SIMBA galaxies, and the other way around.

\subsection{Physical interpretation}

We now discuss the physics behind our results. As we saw in Fig. \ref{fig:Mstar_Vmax_Zstar}, galaxy properties populate differently the parameter space depending on the value of $\Omega_{\rm m}$. This indicates that $\Omega_{\rm m}$ induces an effect on galaxy properties, or on a subset of them, that cannot be mimicked by astrophysical effects. Let's first focus our attention on the two most important properties of both the SIMBA and the IllustrisTNG galaxies: $V_{\rm max}$ and $M_*$. From the top panels of Fig. \ref{fig:Mstar_Vmax_Zstar} we can see that at fixed stellar mass, the maximum circular velocity increases monotonically with $\Omega_{\rm m}$. This may be explained taking into account that higher values of $\Omega_{\rm m}$ will increase the dark matter density in the Universe, and therefore more dark matter is expected to reside in galaxies, enhancing their gravitational potential well and therefore their $V_{\rm max}$ value. However, feedback from supernovae and AGN are also expected to affect the stellar mass of the galaxy, introducing some scatter in the $M_* - V_{\rm max}$ relation. 

As we shall see below, our tests suggest that $\Omega_{\rm m}$ is not imprinted into a single property (e.g. $V_{\rm max}$) and that knowing the value of the astrophysical parameters perfectly does not significantly help. Thus, we may think that $\Omega_{\rm m}$ may change the manifold where galaxy properties live, and that change is different to the one induced by changes in feedback. 

This explanation could shed light on why we cannot determine the value of $\sigma_8$ with a single galaxy. In contrast to $\Omega_{\rm m}$, $\sigma_8$ will only change the amplitude of the initial matter fluctuations, and we think that by itself is unlikely to induce systematic differences in galaxy properties. $\sigma_8$ may however affect the abundance of galaxies, in particular of the most massive ones, similarly as it does for the halo mass function. Thus, while a single galaxy (unless located in the high-mass end) may not be enough to infer $\sigma_8$, a set of galaxies can however be used as a probe of $\sigma_8$. We leave this study for future work.

As we saw in the results section, for some simulations, the predictions of the models seem to exhibit an overall bias. This may be due to the following reason. The networks may be learning some function that approximates the galaxy properties manifold and its dependence with $\Omega_{\rm m}$. However, due to the limited data we have to train them, it may happen that the learned manifold may be off with respect to the true one. In this case we will expect an overall bias between the prediction of the network and the true value for all galaxies in the considered simulation.

\subsection{Breaking degeneracies with astrophysics?}

We may wonder whether the network is aware of the clear dependence on $\Omega_{\rm m}$ of the $V_{\rm max}$ vs $M_*$ relation, but needs additional information to break the degeneracy between cosmology and astrophysics. Maybe in this case the network is using the other properties (e.g. $Z_*$, $R_*$, and ${\rm K}$) to first constrain astrophysics (i.e. feedback parameter values) and then determine cosmology. To test this hypothesis we train a network using as input variables $\{ M_*, V_{\rm max}, A_{\rm SN1}, A_{\rm SN2}, A_{\rm AGN1}, A_{\rm AGN2}\}$. If our hypothesis holds, by providing the network with the true value of the feedback parameters plus $M_*$ and $V_{\rm max}$ it would be able to infer $\Omega_{\rm m}$ accurately. However, we find that this model performs very badly when inferring the value of $\Omega_{\rm m}$: its accuracy decreases by 91\% with respect to the model trained on all variables. This test indicates that the network is not simply extracting information from $V_{\rm max}$ and $M_*$ and using the other variables to break the degeneracies between cosmology and astrophysics.

Next, we test whether knowing the value of the astrophysical parameters adds additional information to the one already contained in the galaxy properties. To quantify this, we train a network using the 17 properties of the galaxies from the IllustrisTNG simulations plus the value of $A_{\rm SN1}, A_{\rm SN2}, A_{\rm AGN1}$, and $A_{\rm AGN2}$. We find that results barely improve: the model accuracy and precision increases by 3\% and 5\%, respectively. This indicates that most of the information the network is extracting is already contained in the internal galaxy properties, and knowing the value of the feedback parameters perfectly does not add any significant additional information. This and the above test indicate that the networks are not trying to infer feedback to break some degeneracies with a particular observable, but rather than the observable itself is sensitive to $\Omega_{\rm m}$ by itself. 

We note that it is well known that galaxy properties change with redshift. This not only explains why the models we train at $z=0$ do not work at higher redshifts, but also why knowing the value of the astrophysical parameters perfectly does not add much information, since these values will be the same across redshifts.

\subsection{Dark matter content}

The explanation we formulated above to interpret our findings relies on dark matter playing a crucial role on galaxies. In order to test this hypothesis, we performed the following test. We have trained networks on galaxies from the IllustrisTNG simulations using all properties except $V_{\rm max}$, $\sigma_v$, $M_{\rm t}$, $R_{\rm t}$, and $R_{\rm max}$. These are quantities that are expected to receive large contributions from the dark matter component of galaxies, and therefore, is a way to quantify how important it is for the network to know the dark matter component or the depth of the gravitational potential well. We find that the network trained with this configuration is still able to infer the value of $\Omega_{\rm m}$ but with much lower accuracy: 96\% worse than the model trained on all properties. This test indicates that these variables are very important, although $\Omega_{\rm m}$ leaves some weaker signatures on the other galaxy properties. In a complementary way, we have checked that in the case of the IllustrisTNG galaxies, once we have identified the 5 most important variables $\{V_{\rm max}, M_*, Z_*, R_*, {\rm K}\}$, removing $V_{\rm max}$ from that set completely cancels the constraining power. In other words, for that subset, $V_{\rm max}$ is needed to infer $\Omega_{\rm m}$. From these tests we conclude that the network may be using information either about the dark matter content of the galaxy or about its gravitational potential well.

\begin{figure*}[t]
\centering
\includegraphics[width=0.99\linewidth]{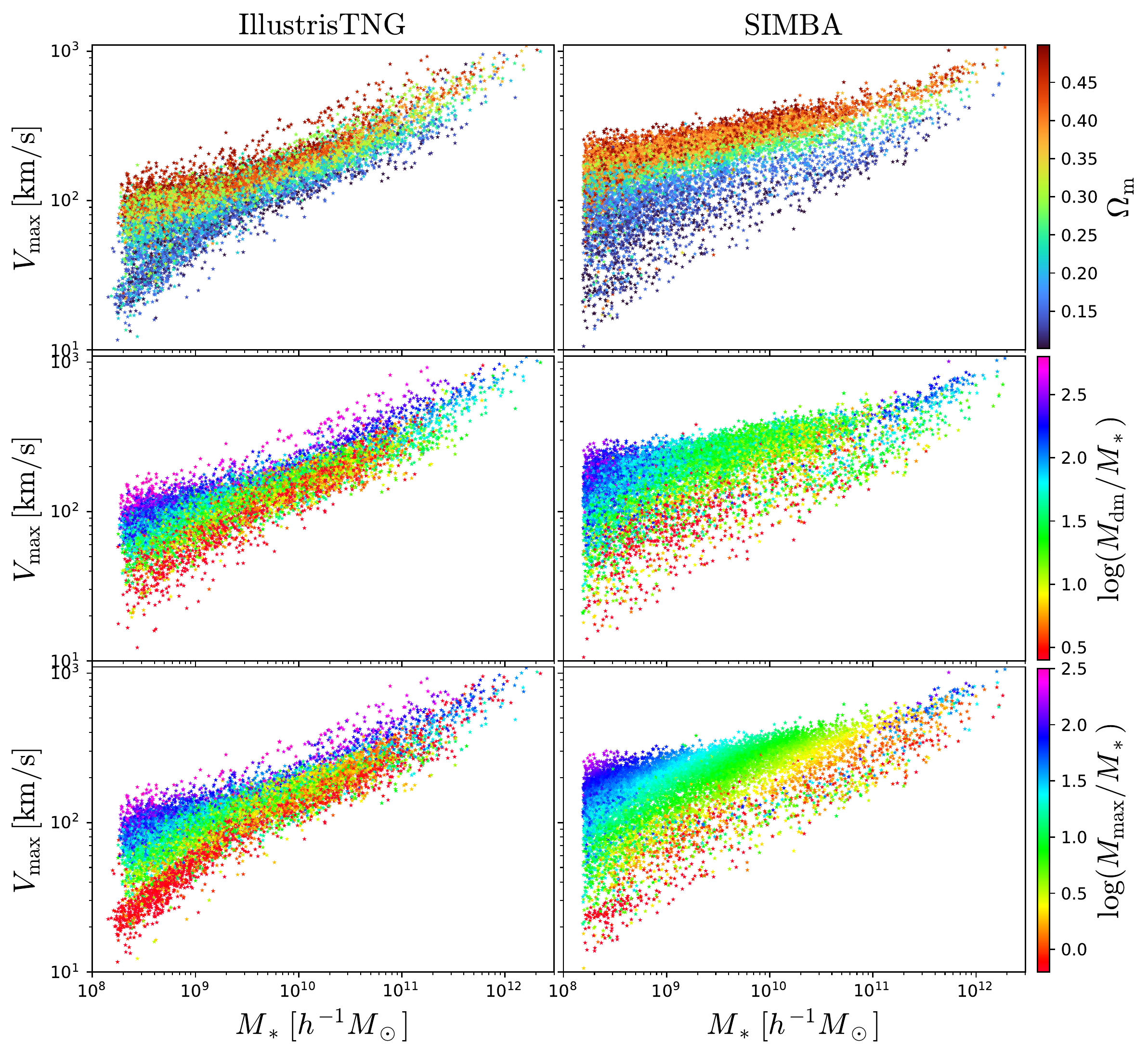}
\caption{We have randomly taken 100 simulations from the IllustrisTNG (left column) and SIMBA (right column). For each simulation we take 100 random galaxies at $z=0$. We then project these galaxies into the $V_{\rm max}-M_*$ plane. Each row shows the data color coded according to the value of $\Omega_{\rm m}$ (top), $\log(M_{\rm dm}/M_*)$ (middle), and $\log(M_{\rm max}/M_*)$ (bottom), where $M_{\rm dm}=M_{\rm t}-M_{\rm g}-M_*-M_{\rm BH}$ is the dark matter mass in the galaxy and $M_{\rm max}=V_{\rm max}^2R_{\rm max}/G$ is the total matter mass contained within $R_{\rm max}$. We find that at fixed stellar mass $V_{\rm max}$ increases with both $\Omega_{\rm m}$ and $M_{\rm dm}$, supporting our hypothesis that increasing the value of $\Omega_{\rm m}$ increases the dark matter content of galaxies, making the gravitational potential deeper and therefore enhancing $V_{\rm max}$. From the third row we can however see that at fixed stellar mass, $V_{\rm max}$ is more strongly correlated with $M_{\rm max}$; this may explain why the network prefers to extract information from $V_{\rm max}$ rather than the subhalo total mass or dark matter mass. We note that the reason why we use $M_{\rm dm}$ and $M_{\rm max}$ normalized to the stellar mass is because there is a strong correlation between these quantities and $M_*$. By taking the ratio we get rid of that dependence, simplifying the visualization of the results.}
\label{fig:Mstar_Vmax_DM}
\end{figure*}

Next, to reinforce our explanation, we test explicitly whether the dark matter content of galaxies increases with $\Omega_{\rm m}$. We have taken the 100 galaxies for 100 different models that we discussed above and plot in Fig. \ref{fig:Mstar_Vmax_DM} the $V_{\rm max}$ versus $M_*$ projection for those galaxies. The top panels show the galaxies color-coded by their value of $\Omega_{\rm m}$, and show the trend we already discussed above (the top panels are identical to the panels in the left column of Fig. \ref{fig:Mstar_Vmax_Zstar}). The panels in the middle row show the results color coded by the ratio between the dark matter mass\footnote{The dark matter mass is computed as $M_{\rm t}-M_{\rm g}-M_*-M_{\rm BH}$.} and stellar mass in the galaxies. We use that ratio and not the dark matter mass as the latter has a strong correlation with stellar mass, making more challenging the visualization. As can be seen, for a fixed value of the stellar mass, the larger the dark matter mass the higher the value of $V_{\rm max}$. This trend is very clear for IllustrisTNG galaxies; meanwhile for SIMBA it is also clear for low- and high-mass galaxies, while for intermediate galaxies ($9.3<\log M_*/(h^{-1}M\odot) < 11$) the dependence is much weaker. This is the same trend we find with $\Omega_{\rm m}$ (top panels), indicating that larger values of $\Omega_{\rm m}$ will tend to increase the dark matter content of galaxies.

We note that increasing the dark matter content of galaxies can also affect other galaxy properties. For instance, changing $\Omega_{\rm m}$ will affect halo collapse time and concentration, and these may leave an imprint on $Z_*$, $R_{\rm max}$ and $R_*$. However, the relationship between these variables and the $V_{\rm max} - M_*$ plane is not clearly visualized in a 3-dimensional plot (two plus color) as in Fig. \ref{fig:Mstar_Vmax_DM}.  We argue that this is due to the high-dimensional manifold on which these features depend on $\Omega_{\rm m}$. 

On the other hand, we may also expect that differences in $\sigma_8$ will led to changes in halo formation time and concentration. Since we cannot infer the value of $\sigma_8$ from individual properties of galaxies, we think the effect of $\Omega_{\rm m}$ on galaxy properties should not be primarily driven by the above changes to halo properties; or perhaps a distinct change to the one induced by $\sigma_8$.

\subsection{$V_{\rm max}$ vs $M_{\rm t}$}

The above results corroborate our interpretation that changing the value of $\Omega_{\rm m}$ affects the dark matter content of galaxies; an effect that is physically different to the one from feedback. However, at this point we may wonder why the network prefers to use $V_{\rm max}$ rather than other properties that are expected to be heavily affected by dark matter such as the galaxy's subhalo total mass ($M_{\rm tot}$) or velocity dispersion ($\sigma_v$). To verify that this is indeed the case, we have trained models with galaxies of the IllustrisTNG simulations using as features $\{ M_*, M_{\rm t}, Z_*, R_*, {\rm K}\}$ and $\{ M_*, \sigma_v, Z_*, R_*, {\rm K}\}$. We find that using these variables the accuracy of the model on $\Omega_{\rm m}$ degrades by 100\% and 43\%, respectively. This clearly indicates that $V_{\rm max}$ contains more information than $M_{\rm t}$ and $\sigma_v$. We believe that this may be happening because it is known that $V_{\rm max}$ correlates more strongly with stellar mass than with subhalo mass \citep{Conroy_2006}. For instance, when halos are accreted into larger halos they may lose a significant fraction of their dark matter content due to tidal forces. That effect will change the dark matter content of galaxies significantly, but the value of $V_{\rm max}$ may remain rather stable since it mostly probes the mass in the inner regions of the subhalo, that are the least affected by the above processes. 

To validate this hypothesis we plot in the bottom row of Fig. \ref{fig:Mstar_Vmax_DM} the galaxies mentioned above but color coded according to $M_{\rm max}/M_*$, where $M_{\rm max}=V_{\rm max}^2R_{\rm max}/G$. We find for IllustrisTNG galaxies a similar trend as when we used the dark matter mass, while for SIMBA galaxies the trend is now much more evident: for a fixed stellar mass, increasing the value of $M_{\rm max}$ increases the value of $V_{\rm max}$. This indicates that either $V_{\rm max}$ (or $M_{\rm max}$) is a better and more stable proxy for the dark matter content of galaxies than the total subhalo mass or its velocity dispersion. 

The above tests may indicate that the network is focusing its attention on the dark matter or total mass content of galaxies in their central region, or maybe directly into the depth of the gravitational potential, rather than in the total dark matter mass in the subhalo's galaxy.

\section{Summary \& Discussion}
\label{sec:summary}

In this paper we have shown that it may be possible to infer the value of $\Omega_{\rm m}$ with a precision of $\delta \Omega_{\rm m}/\Omega_{\rm m}\simeq10-15\%$ and an accuracy of $\sim0.035 - 0.042$ from the internal properties of individual galaxies and their subhalos. This result holds for galaxies of either the CAMELS-IllustrisTNG and CAMELS-SIMBA simulations, when using neural networks (to do likelihood-free inference) or gradient boosting trees (to do parameter regression), and at all redshifts considered $z \leq 3$. 

We have shown that $\Omega_{\rm m}$ has a large effect on the $V_{\rm max}-M_*$ relation, although our constraints do not arise from those two variables alone, even if the astrophysical parameters are known perfectly. We believe that the explanation behind our results is that galaxy properties reside in a high-dimensional manifold that changes with $\Omega_{\rm m}$. That change is different to the one induced by astrophysical effects. We think that the physics behind the unique change in the manifold is that $\Omega_{\rm m}$ affects the dark matter content of galaxies. Machine learning methods can be trained to find these manifolds and therefore to infer the value of $\Omega_{\rm m}$. 

We note that physically, the effect of $\Omega_{\rm m}$ is very different to the one of $\sigma_8$, which will just change the amplitude of the initial linear matter fluctuations and therefore we do not expect it to imprint unique features on galaxy properties. This could explain why our models cannot infer the value of $\sigma_8$ from individual galaxy properties.

\subsection{Robustness}

We caution the reader that our models are not robust; if the models are trained on galaxies from the IllustrisTNG simulations, they cannot infer the value of $\Omega_{\rm m}$ from galaxies of the SIMBA simulations, and vice versa. We believe that this may be due to the intrinsic differences between the galaxy properties in the two different simulations (see Fig. \ref{fig:Mstar_Vmax_Zstar}).

\begin{figure*}
\centering
\includegraphics[width=0.99\linewidth]{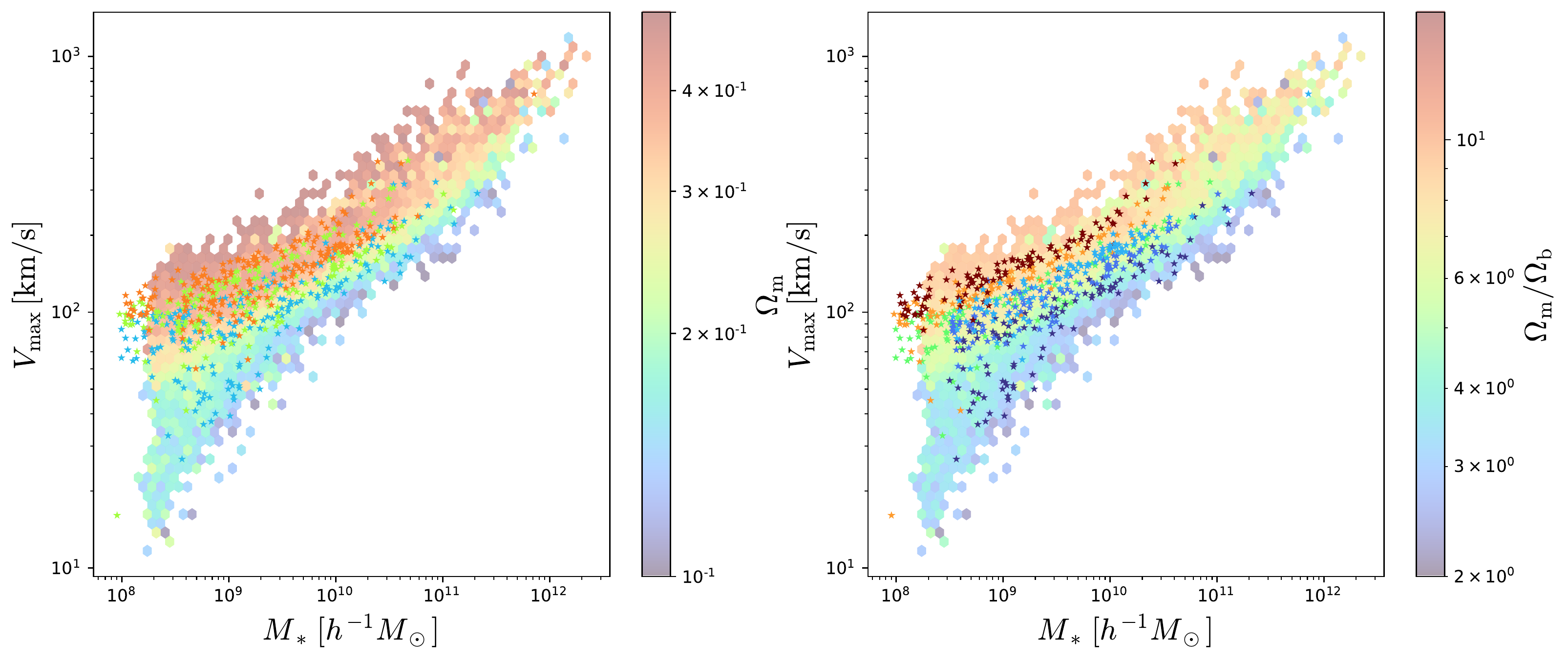}
\caption{In order to explore in a very qualitative manner whether our method is sensitive to $\Omega_{\rm b}/\Omega_{\rm m}$ or to $\Omega_{\rm m}$ and/or $\Omega_{\rm b}$ we have run a set of six simulations with different values of $\Omega_{\rm b}$ (0.025 and 0.075) and $\Omega_{\rm m}$ (0.2, 0.3, and 0.4) using the AREPO and IllustrisTNG model (using the fiducial astrophysical model). For each of those simulations we have randomly taken 100 galaxies. In the two different panels we show $V_{\rm max}$ versus $M_*$ of those galaxies color coded according to their value of $\Omega_{\rm m}$ (left) and $\Omega_{\rm m}/\Omega_{\rm b}$ (right). In the background we show with a hexbin plot the distribution of galaxies from the IllustrisTNG simulations. From the left panel we can clearly see that galaxies no longer follow a monotonic relation of increasing $V_{\rm max}$ with $\Omega_{\rm m}$. On the other hand, from the right panel we can see a much more steady and monotonic relation when using $\Omega_{\rm m}/\Omega_{\rm b}$. We however note that the colors of the galaxies do not really match the ones from the background simulations with fixed $\Omega_{\rm b}$.}
\label{fig:Omega_b}
\end{figure*}

While this method, in its current form, cannot be used with real data yet due to the lack of robustness, it will be interesting to explore the use of contrastive learning \citep{Contrastive_learning} to force the network to learn only unique (physical) features that are not simulation/model dependent. Another possible avenue will be to try to develop a theoretical template (e.g. using symbolic regression) and calibrate its parameters directly with real data. We leave these questions for future works. 

Once the model is robust, it will be important to quantify how much our constraints degrade by accounting for the observational uncertainties associated to the different galaxy properties. On the other hand, if our interpretation is correct and galaxy properties live in a manifold sensitive to cosmology and astrophysics, one can use that information to reduce uncertainties in galaxy properties by requiring them to be in a manifold. In other words, in the real Universe, galaxy properties will reside in a manifold with a fixed cosmology and astrophysics. Thus, there will be high-dimensional correlations that may allow us to determine the value of some galaxy properties with higher accuracy.

\subsection{$\Omega_{\rm b}$}

Due to the design of the CAMELS simulations, we can only train models to infer the value of $\Omega_{\rm m}$ and $\sigma_8$, since in all simulations we have kept fixed the value of the other cosmological parameters. It will be important to repeat this work using simulations that vary the value of other cosmological parameters to investigate whether individual galaxies can constrain other parameters but also to study whether degeneracies among parameters will deteriorate the constraining power of this method on $\Omega_{\rm m}$. 

In Sec. \ref{sec:interpretation} we have seen that our models rely on both galaxy properties and the depth of the gravitational potential well (or the mass in the galaxy core) to infer the value of $\Omega_{\rm m}$. Thus, galaxy properties may also be sensitive to $\Omega_{\rm b}$, as varying that parameter will change the abundance of baryons in galaxies. Thus, it will be interesting to investigate whether galaxy properties are also sensitive to $\Omega_{\rm b}$. On the other hand, it may happen that galaxy properties are sensitive to some particular combination of $\Omega_{\rm b}$ and $\Omega_{\rm m}$, e.g. to its ratio: $\Omega_{\rm b}/\Omega_{\rm m}$. While we cannot provide an answer to these questions (as it will require running many simulations with different values of $\Omega_{\rm m}$) we can however attempt to provide a qualitative indication of what may be happening. For this, we have run 6 additional IllustrisTNG simulations. In these simulations the value of the astrophysical parameters is set to the fiducial IllustrisTNG model, while $\sigma_8$ is 0.8 and $\{\Omega_{\rm m}, \Omega_{\rm b}\}$ is given by $\{0.2, 0.025\}$, $\{ 0.2, 0.075\}$, $\{0.3, 0.025\}$, $\{ 0.3, 0.075\}$, $\{0.4, 0.025\}$, $\{ 0.4, 0.075\}$. For each of those 6 simulations we randomly select one hundred galaxies. 

In Fig. \ref{fig:Omega_b} we show these galaxies projected in the $V_{\rm max}-M_*$ plane. Galaxies are color coded according to the value of $\Omega_{\rm m}$ (left) and $\Omega_{\rm m}/\Omega_{\rm b}$ (right). In the background we show a hexbin plot with the distribution of galaxies from the 1,000 IllustrisTNG simulations with fixed value of $\Omega_{\rm b}$. As can be seen, for a fixed value of $M_*$, galaxies do not follow a monotonic relation of higher $V_{\rm max}$ for larger $\Omega_{\rm m}$. It seems that in this case the two different values of $\Omega_{\rm b}$ create a bimodal distribution. 

In the right panel of Fig. \ref{fig:Omega_b} we color code the same galaxies as before but using instead $\Omega_{\rm m}/\Omega_{\rm b}$. In this case, we find a more monotonic relation between $V_{\rm max}$ and $\Omega_{\rm m}/\Omega_{\rm b}$ at fixed stellar mass. We note however that the colors of these galaxies are a bit off with respect to the ones from the IllustrisTNG set with fixed $\Omega_{\rm b}$. Thus, while these results indicate that $\Omega_{\rm b}/\Omega_{\rm m}$ is a more relevant variable than $\Omega_{\rm m}$ when $\Omega_{\rm b}$ is not fixed, we cannot tell whether our method is just sensible to $\Omega_{\rm m}/\Omega_{\rm b}$ or whether in higher dimensions degeneracies can be broken and we can constrain both $\Omega_{\rm m}$ and $\Omega_{\rm b}$. We note that the value of $\Omega_{\rm b}/\Omega_{\rm m}$ can be constrained from cosmic microwave background data with high accuracy. Thus, if galaxy properties are indeed sensitive to $\Omega_{\rm b}/\Omega_{\rm m}$, it will be a interesting way to connect two very different observables and physical quantities of the Universe.

We emphasize that we have not provided a full physical interpretation of the results presented in this work, beyond stating that changing $\Omega_{\rm m}$ affects galaxy properties in a way different to the one produced by changing astrophysics parameters. However, we know that the dark matter content/total matter content/depth of the gravitational potential is a very important variable for the network. Besides, our results indicate that the network may be more sensitive to $\Omega_{\rm b}/\Omega_{\rm m}$ rather than $\Omega_{\rm m}$. One may wonder if the network is somehow measuring the total mass in the center of the galaxy (e.g. through the depth of the gravitational potential well) and also measuring the mass in baryons in that region. This would allow the network to directly infer $\Omega_{\rm b}/\Omega_{\rm m}$. In the future, it may be interesting to explore whether there are relations between the total matter and the baryonic content in the inner regions of galaxies that somehow are robust to changes in astrophysics. We note that the idea of measuring $\Omega_{\rm b}/\Omega_{\rm m}$ from individual objects was outlined in \citet{White_1993}, although there it only applied to the most massive halos where feedback effects cannot expel baryons out to the intergalactic medium.

\subsection{Numerical effects}

Given the surprising results our models have achieved, we should ask ourselves: where does the information come from? In other words, is the network extracting information from a physical or a numerical effect? 

$\Omega_{\rm m}$ is imprinted in the simulations through several different effects; for example: 1) it will affect the amplitude and shape of the linear matter power spectrum used to generate the initial conditions, 2) it will affect the mass of the dark matter particles, and 3) it will change the expansion rate. 

If the networks are using some non-physical feature to get the value of $\Omega_{\rm m}$ from the changes to the power spectrum, it would be expected that they would also be able to infer the value of $\sigma_8$ that also affects the linear power spectrum. Since our models are unable to constrain the value of $\sigma_8$, we believe this effect should not be the cause of our results.

The one-to-one correlation between $\Omega_{\rm m}$ and the masses of the dark matter particles (note that in CAMELS $\Omega_{\rm b}$ is kept fixed at 0.049 in all simulations) is something that can be easily learned by neural networks, but is not a physical effect. However, in the considered galaxy properties there is no obvious way where this effect can show up. The dark matter mass of subhalos obeys the relation $M_{\rm DM}=N_{\rm dm}m_{\rm dm}$, where $m_{\rm dm}$ is the mass of a dark matter particle and $N_{\rm dm}$ is the number of dark matter particles in the subhalo, which should be an integer number. There is an intrinsic degeneracy between $N_{\rm dm}$ and $m_{\rm dm}$ for this to work. Besides, if this would be the case, the network would need to estimate the mass in dark matter of the halo by subtracting the gas, stellar, and black-hole mass to the total mass of the subhalo. We know from our analysis of the relevant features that none of those properties are important for the networks. Other properties, like $V_{\rm max}$, $\sigma_v$, $V$, $R_t$, and $R_{\rm max}$ seem more unlikely to be easily related to the masses of the dark matter particles. Thus, we find this hypothesis not very likely.

$\Omega_{\rm m}$ will also change the expansion rate history in the simulation. However, we cannot think of a situation where the model may be learning a numerical artifact associated to this.

Finally, we note that $\Omega_{\rm b}/\Omega_{\rm m}$ is important to set the internal structure of galaxies (e.g. how baryon dominated the rotation curve is). Thus, the density of gas in the galaxy is expected to be affected by this, which in turn will affect cooling and feedback. However, these effects are highly non-linear and is not obvious whether numerical effects can be imprinted on them.

Thus, while we could not identify a process that will give rise to a numerical artifact that can be learned by the machine learning models, we cannot completely discard that possibility here.

\subsection{Linear information}
\label{sec:linear}

On average, our models are able to constrain the value of $\Omega_{\rm m}$ with a $\sim10\%$ precision and an accuracy of $\sim 0.03$ for a single, generic galaxy. We may wonder whether there is enough modes in the Lagrangian region of those galaxies to achieve such accuracy. To provide an answer to this question we consider a volume $V$ and use the Fisher matrix formalism to quantify how much information that volume contains, considering it probes scales from $k_{\rm min}\sim2\pi/V^{1/3}$ to $k_{\rm max}=64~h{\rm Mpc}^{-1}$. The value of $k_{\rm max}$ arises from the Nyquist frequency used to generate the initial conditions.

\begin{figure}[t]
\centering
\includegraphics[width=0.99\linewidth]{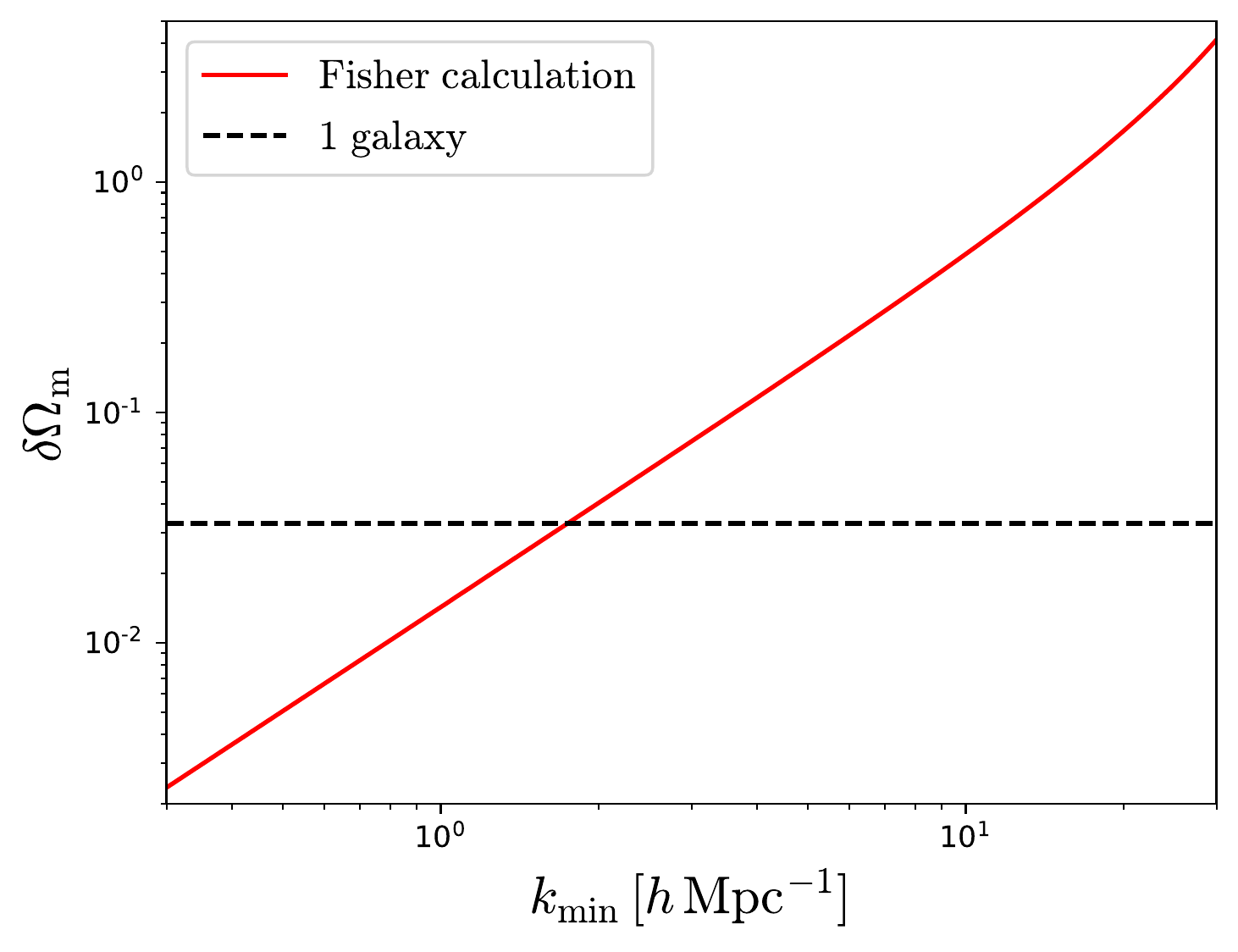}
\caption{We have used the Fisher matrix formalism to calculate how much information there is in the linear, Gaussian density field, for a cosmological volume $V$ considering it contains modes from $k_{\rm min}=2\pi/V^{1/3}$ to $k_{\rm max}=64~h{\rm Mpc}^{-1}$. The solid red line shows the constraints on $\Omega_{\rm m}$ as a function of $k_{\rm min}$, while the dashed black line displays the average error on $\Omega_{\rm m}$ from our models. We can see that only volumes larger than $V\sim(3~h^{-1}{\rm Mpc})^3$ will contain enough modes to be able to place a constraint on $\Omega_{\rm m}$ similar, or better, to the one we obtain. We expect this volume to be larger than the Lagrangian region of most galaxies considered.}
\label{fig:linear_information}
\end{figure}

For our setup, the Fisher matrix can be computed as
\begin{equation}
F_{\alpha \beta}=\int_{k_{\rm min}}^{k_{\rm max}} V\frac{\partial \log P(k,\vec{\theta})}{\partial \theta_\alpha}\frac{\partial \log P(k,\vec{\theta})}{\partial \theta_\beta}\frac{k^2dk}{(2\pi)^2}
\end{equation}
where in our case $\vec{\theta}=\{\Omega_{\rm m}, \sigma_8\}$, $P(k,\vec{\theta})$ is the linear matter power spectrum, and $V$ the cosmological volume. The integral goes from $k_{\rm min}\sim2\pi/V^{1/3}$ to $k_{\rm max}$.

In Fig. \ref{fig:linear_information} we show with a red solid line the marginalized constraints on $\Omega_{\rm m}$ as a function of $k_{\rm min}$. As can be seen, to achieve an error on $\Omega_{\rm m}$ below 0.033 we need a value of $k_{\rm min}\sim2~h{\rm Mpc}^{-1}$, or a Lagrangian region of volume $\sim (\pi~h^{-1}{\rm Mpc})^3$. 

We expect the Lagrangian volume of most galaxies to be smaller than the above estimate \citep[see e.g.][]{Onorbe_2014}, indicating that the constraints from our models are better than the ones that can be obtained from a linear Gaussian field of the same volume. However, there are several caveats to this calculation. First of all, on scales smaller than $\sim1~h{{\rm Mpc}}^{-1}$ the non-linear matter density field may contain information not contained in the initial Gaussian field \citep{Bayer_2021b}. Besides, our models include properties related to velocities (e.g. the galaxy peculiar velocity or the subhalo velocity dispersion) that can also provide additional information to the one contained at linear order. On top of this, on very small scales cosmological modes are expected to be tightly coupled. Thus, even a relatively small Lagrangian region of a galaxy may be affected by modes larger than it, that could add additional information. 

From this test we cannot draw any definitive conclusion on whether the constraints from our models are physical or they just reflect some nonphysical information arising from numerical artifacts. 

\subsection{Consequences}
\label{subsec:consequences}

Our results suggest that galaxy properties will reside in different manifolds for different values of $\Omega_{\rm m}$. This in turn implies that it should be difficult, if not impossible, to reproduce the galaxy properties from real galaxies for cosmologies with a value of $\Omega_{\rm m}$ far away from the true one. This is a clear prediction of this work that can be tested either using hydrodynamical simulations or semi-analytic models.

Regarding hydrodynamic simulations, in CAMELS we vary four astrophysical parameters, while many others are kept fixed. In order to claim that $\Omega_{\rm m}$ induces a distinct effect on galaxy properties it is important to repeat the analyses carried out in this paper but sampling a much larger volume in parameter space where all astrophysical parameters are varied. This will allow us to investigate whether other astrophysical parameters may mimic the effect of $\Omega_{\rm m}$ on galaxy properties.

On a side note, we note that galaxy properties are known to exhibit some level of intrinsic stochasticity \citep{Butterfly} in numerical simulations. If our interpretation of the results is correct, this will imply that either the manifold containing the galaxy properties will have some intrinsic tightness, or that galaxies affected by this effect will move along the manifold.

\subsection{Future work}

In this work we have focused our attention on individual galaxies. In future work we will investigate the improvement on the parameter constraints when considering several galaxies instead of just one. We think that in this case the manifold where galaxies reside will be much better constrained and therefore we expect tighter constraints on all parameters. Furthermore, with many galaxies it may be possible to extract information from different summary statistics (e.g. stellar mass function) that may not be contained in the above manifolds.

While in this paper we have focused our attention on inferring the value of $\Omega_{\rm m}$ from individual galaxies, in the appendix \ref{sec:astro_params} we show that this method can also be used to infer the value of some astrophysical parameters. Given the accurate measurements of the value of the cosmological parameters from other methods, we may consider that this method may be used as a direct probe of astrophysical effects by fixing the value of the cosmological parameters. We will explore this direction in future work. 

Most of the properties considered in this work can be measured from surveys. However, some of them, like the maximum circular velocity and the velocity dispersion, cannot be easily measured for large galaxy populations. In future work we plan to substitute those properties by others that can be measured, e.g. velocity dispersion of the stars or neutral hydrogen, and investigate whether the value of $\Omega_{\rm m}$ can still be inferred with them. It will be also interesting to quantify the accuracy gained by adding other properties not used in this work such as galaxy morphology, stellar age, mass in neutral, molecular, and different individual metals, and environmental quantities like the overdensity of matter or galaxies in a given scale. Given the similarities we have outlined in this work with subhalo abundance matching\footnote{For instance, our results suggest that $V_{\rm max}$ and $M_*$ are the most important variables. These variables play a crucial role in subhalo abundance matching.}, it will be worth investigating whether our results improve even for non observational quantities like that peak of the maximum circular velocity that are known to be better correlated to stellar mass than $V_{\rm max}$. 

Further work is also needed to investigate the dependence of our results on resolution, i.e. whether the manifold where galaxy properties live also changes with the mass and spatial resolution of the simulations and on the algorithm used to identify halos, subhalos, and galaxies. We will also investigate whether our results still hold when using semi-analytic models instead of hydrodynamic simulations; we will use CAMELS-SAM \citep{Lucia_2022} for this.

Another important avenue to take in this work is the use of more interpretable machine learning techniques, such as symbolic regression. These techniques are designed to provide analytic expressions between sets of variables, and their functional form may be easier to interpret than neural networks and gradient boosting trees. We note that we have used such techniques in this work but we were not able to obtain expressions accurate enough to capture the underlying relation. We thus leave this research direction for future work.

We believe that this work illustrates the complex interplay between cosmology and astrophysics on different physical scales (from galactic to cosmological) and how cosmological information may still be present within objects shaped by complex astrophysical processes such as galaxies. We also think this work shows how the use of machine learning techniques can help us better understand and disentangle complex physical processes and discover new features and techniques to maximize the information we can extract from the data.

In order to enable the community to reproduce our results, and to perform additional tests not carried out in this work we made all data and codes used in this work publicly available. We refer the reader to \url{https://github.com/franciscovillaescusa/Cosmo1gal} for further details.

\section*{ACKNOWLEDGEMENTS}
We thank Tom Abel, Arka Banerjee, Adrian Bayer, Greg Bryan, Neal Dalal, ChangHoon Hahn, Andrew Hearin, Lars Hernquist, Oliver Philcox, Tjitske Starkenburg, Michael Strauss, Masahiro Takada, and Benjamin Wandelt for useful conversations. We thank Uros Seljak for suggesting us to perform the calculation of Sec. \ref{sec:linear} and Volker Springel for correspondence that gave rise to Sec \ref{subsec:consequences}. We have made use of the XGB\footnote{\url{https://xgboost.readthedocs.io}} and SHAP\footnote{\url{https://shap.readthedocs.io}} packages. The neural networks have been trained using GPUs at the Tiger cluster at Princeton University and the Rusty cluster of the Flatiron Institute. The work of FVN is supported by the Simons Foundation. DAA was supported in part by NSF grants AST-2009687 and AST-2108944. CH is funded by the Deutsche Forschungsgemeinschaft (DFG, German Research Foundation) under Germany’s Excellence Strategy EXC 2121 Quantum Universe-390833306. All the data and codes used for this work are publicly available in \url{https://github.com/franciscovillaescusa/Cosmo1gal}. Details on the CAMELS simulations can be found in \url{https://www.camel-simulations.org}.

\appendix 

\section{Results for SIMBA galaxies}
\label{sec:SIMBA}

In order to verify that our results hold for both IllustrisTNG and SIMBA galaxies, we have repeated the exercise of Sec. \ref{sec:results} and trained neural networks on individual properties of SIMBA galaxies to infer the value of the cosmological and astrophysical parameters.

\begin{figure*}
\centering
\includegraphics[width=0.99\linewidth]{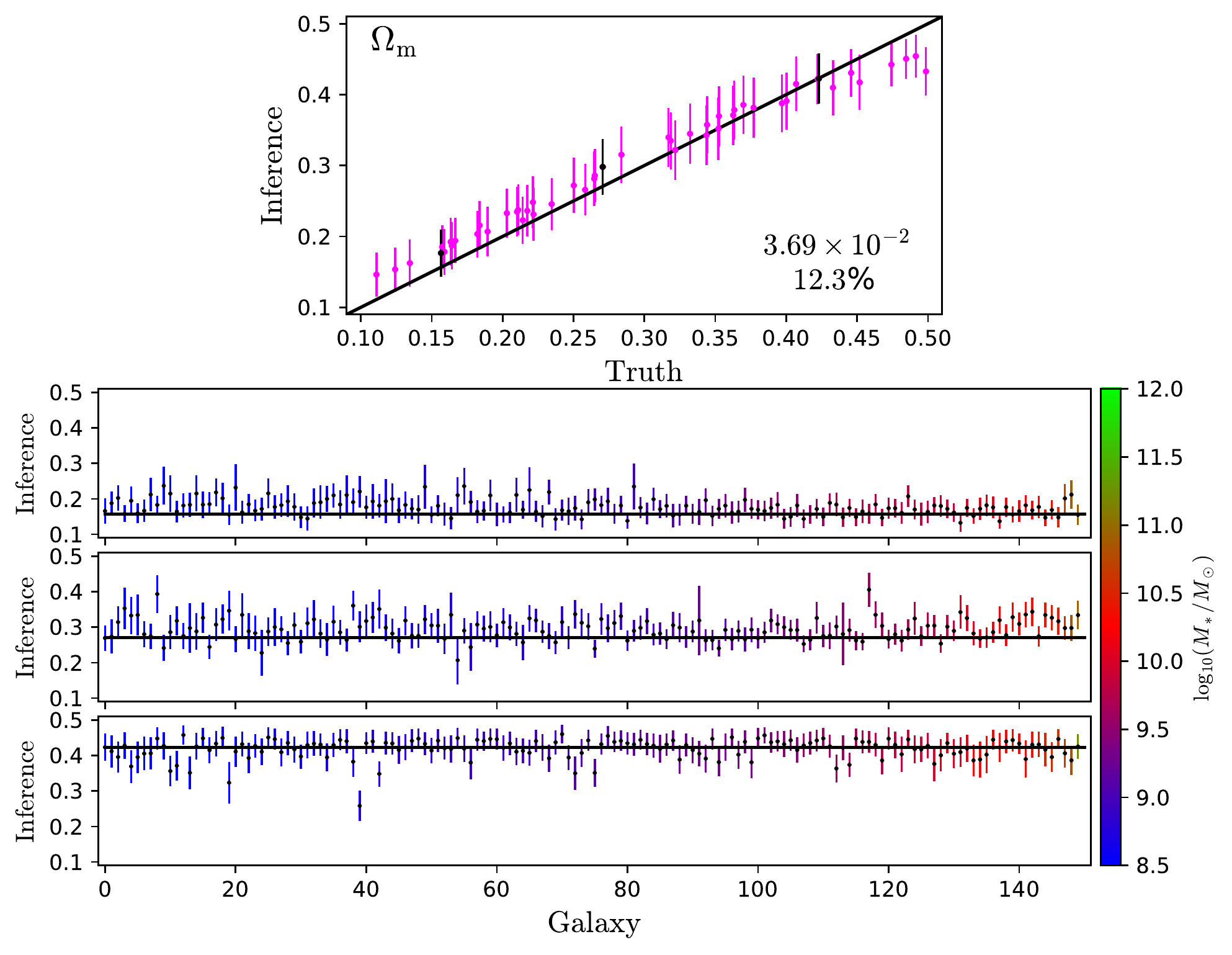}
\caption{Same as Fig. \ref{fig:IllustrisTNG_big} but for SIMBA galaxies at $z=0$.}
\label{fig:SIMBA_big}
\end{figure*}

We show the results in Fig. \ref{fig:SIMBA_big}. We find that, qualitatively, the results for SIMBA galaxies are the same as for IllustrisTNG galaxies. The model is able to infer the value of $\Omega_{\rm m}$ with an accuracy of $\sim3.7\times10^{-2}$ and a precision of $12\%$. We note that we observe a generic bias for true values of $\Omega_{\rm m}$ below $\sim0.35$. This bias seems to be more severe for SIMBA galaxies than for IllustrisTNG galaxies, even when training on 14 variables (see Fig. \ref{fig:robustness}). 

From the bottom panels of Fig. \ref{fig:SIMBA_big} we can see that the network works for any generic galaxy, not a subset of them. As in the case of IllustrisTNG galaxies, we find a very small fraction of outliers. While the precision of the model when inferring $\Omega_{\rm m}$ is very similar for all galaxies when the true value of $\Omega_{\rm m}$ is intermediate or high, we find that the model is more precise when using massive galaxies of models with low values of $\Omega_{\rm m}$. This is similar to what found for IllustrisTNG galaxies, although in that case the differences were even higher.

Overall, we conclude that we can use machine learning methods to constrain the value of $\Omega_{\rm m}$ independently of the simulation suite used to train the model. We emphasize however that our models are not robust (see Sec. \ref{subsec:robust}).

\section{Robustness test}
\label{sec:appendix_robustness}

In Sec. \ref{subsec:robust} we investigated the robustness of our models, finding that they are not robust. In other words, training the models on galaxies from one simulation suite does not allow to infer the correct value of $\Omega_{\rm m}$ from galaxies of other the other simulation suite. In this appendix we show a few more details on this test, investigating whether this is a generic feature for all galaxies or whether the model works in some cases.

\begin{figure*}
\centering
\includegraphics[width=0.99\linewidth]{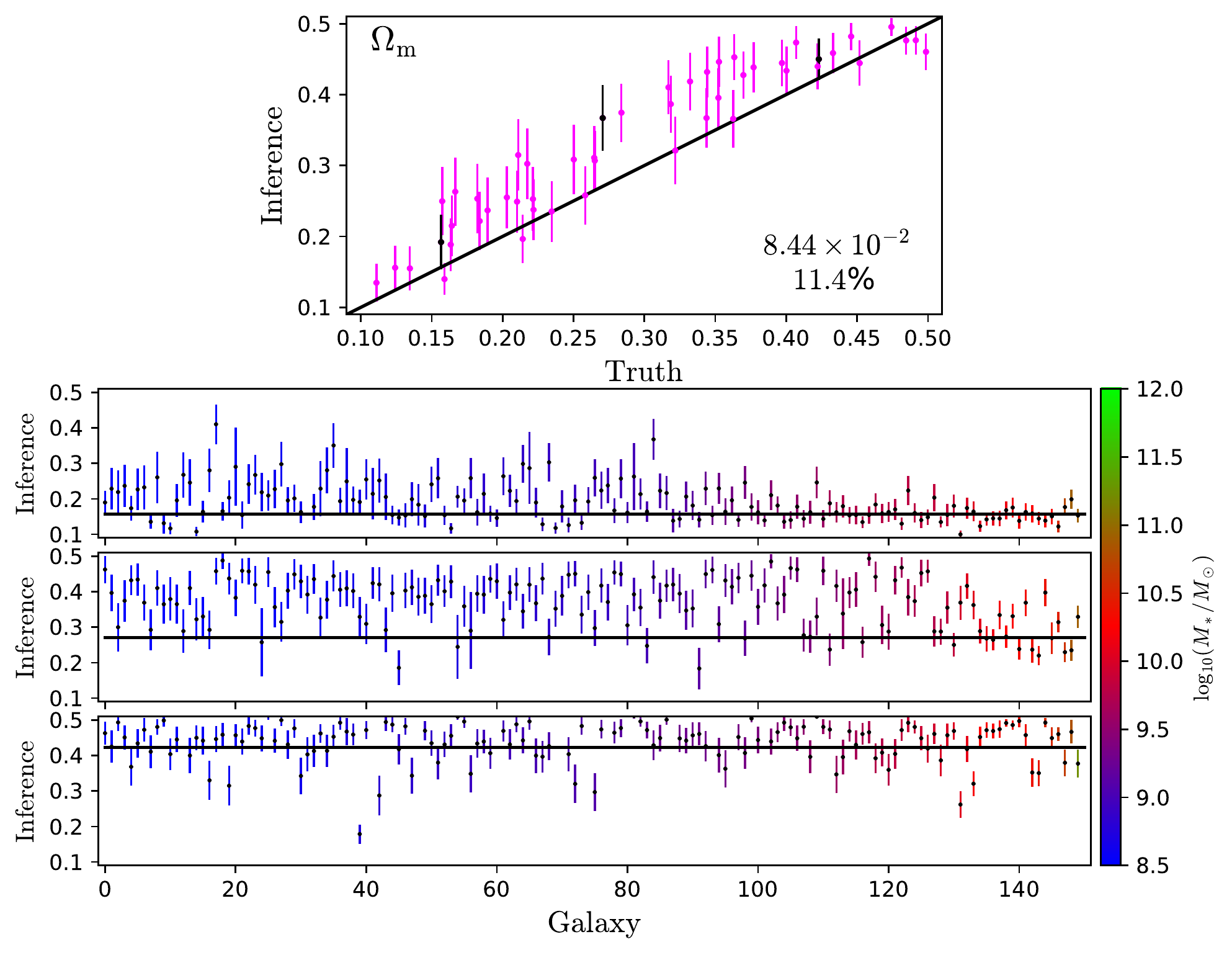}
\caption{Same as Fig. \ref{fig:IllustrisTNG_big} but for a model trained on IllustrisTNG galaxies and tested on SIMBA galaxies.}
\label{fig:robust_big}
\end{figure*}

We have trained a model using galaxies from the IllustrisTNG simulations at $z=0$ (using all properties except the magnitudes in the U, K, and g bands) and tested it on individual galaxies of the SIMBA simulations. In Fig. \ref{fig:robust_big} we show the results of performing the detailed analysis outlined in Sec. \ref{sec:results}. As we already saw in Fig. \ref{fig:robustness} we find that on average, the model is not able to infer the correct value of $\Omega_{\rm m}$ (top panel). We however perform a more detailed analysis of many individual galaxies and show the results in the bottom panels of Fig. \ref{fig:robust_big}. As can be seen, in general, the model does not work for a generic galaxy. On the other hand, results are not completely off; for instance see Fig. 4 of \citet{Paco_2021a} for a similar exercise with 2D maps. We find that the true value of $\Omega_{\rm m}$ lies within the model standard deviation in a large fraction of galaxies, although there is obviously a large underlying bias.

We note that the model works better for cosmologies with low and high values of $\Omega_{\rm m}$ and performs worse for intermediate values. However, this may just an artifact: e.g. the network may be using information from priors. For the model with a true value of $\Omega_{\rm m}\sim0.27$ there is still a non negligible fraction of galaxies where the model seems to eb working. This does not look like the fraction of outliers we have seen in all models in the main text. We defer to future work the exploration of the properties of these galaxies and whether they exhibit more similarities with the ones from the IllustrisTNG simulations.

\section{SHAP values}
\label{sec:SHAP}

\begin{figure*}
\centering
\includegraphics[width=0.49\linewidth]{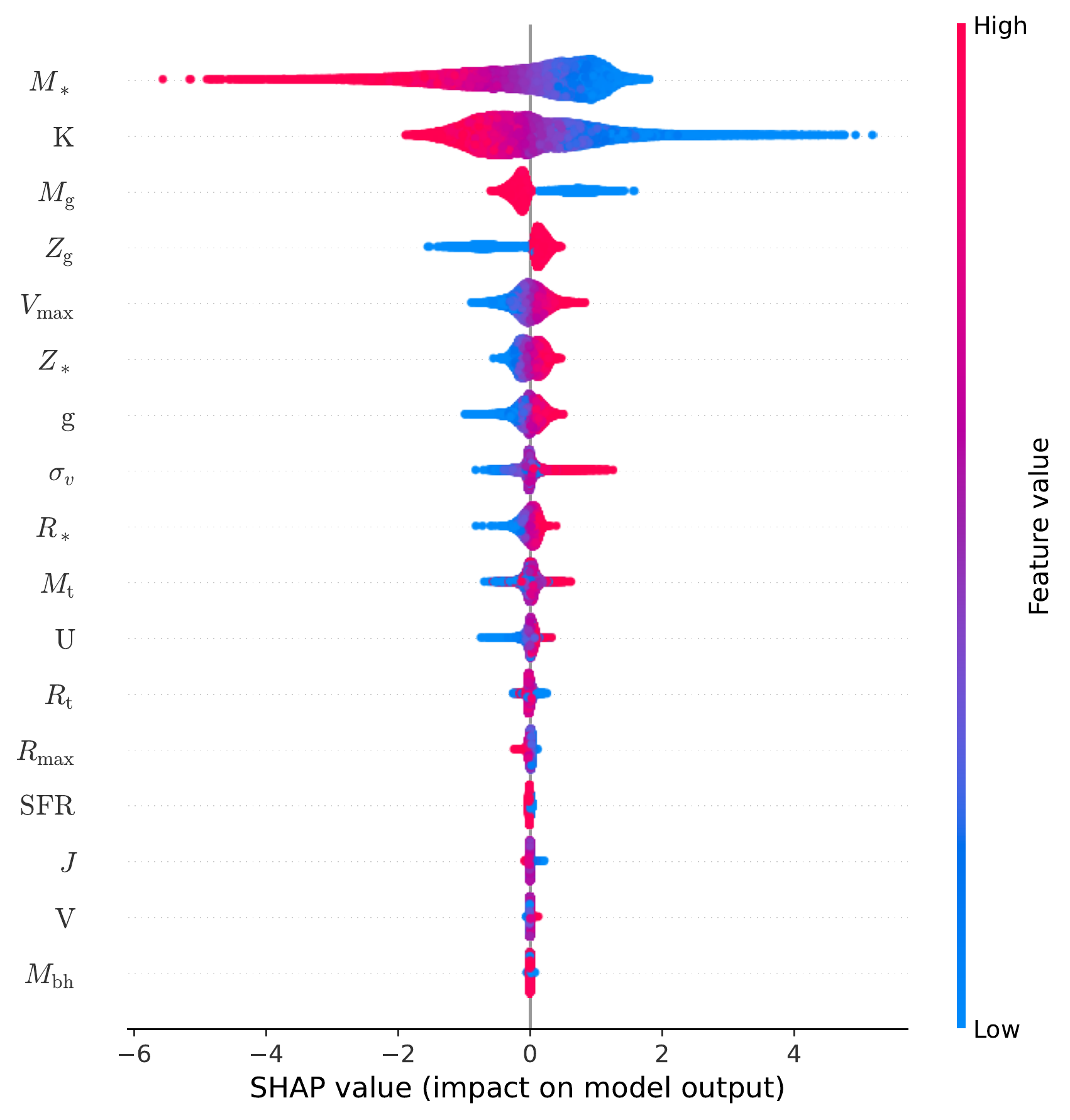}
\includegraphics[width=0.49\linewidth]{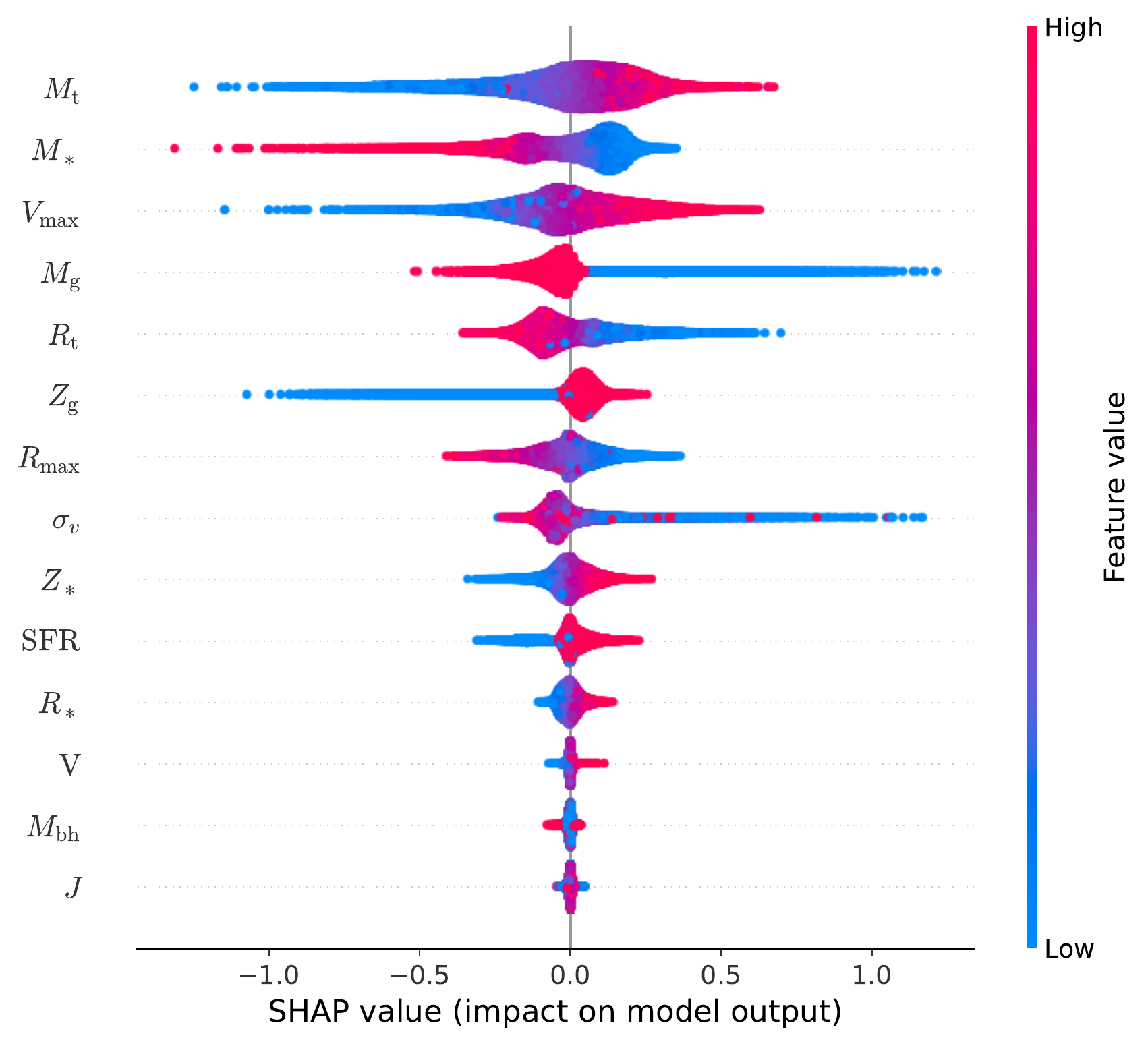}
\caption{In order to identify the most important variables used by the model in order to carry out its predictions we have computed the SHAP (SHapley Additive exPlanation) values for each galaxy in the test set. The panels show the distribution of SHAP values for the galaxies of IllustrisTNG (left) and SIMBA (right) simulations sorted by the different features. The color indicates the value of the variable from low (blue) to high (red). Larger absolute values indicate that the considered feature has a larger impact on the model final prediction.}
\label{fig:shap}
\end{figure*}

In order to identify the most important features of our networks we have computed the SHAP (SHapley Additive exPlanation) value of each galaxy property. This method assigns to each feature of each galaxy a value; larger absolute values for a given property indicates that the feature is having a larger contribution to the final output of the model. In Fig. \ref{fig:shap} we show the distribution of SHAP values for the different features for the models trained on IllustrisTNG (left) and SIMBA (right) galaxies.

For the IllustrisTNG simulations we find that features such as stellar mass, K band, gas mass, gas metallicity, and maximum circular velocity to be among the most important variables. For SIMBA instead we get properties like total mass, stellar mass, maximum circular velocity, gas mass, and subhalo radius. In order to determine whether these variables are indeed the most important ones we have retrained neural networks using as input those five variables instead of the 17/14 original ones from IllustrisTNG/SIMBA. However, the performance of the models trained on these variables is relatively poor; much worse than the variables identified in Sec. \ref{subsec:features}. We think that the reason behind this is that there are multiple variables that are highly correlated, and the model may be extracting information from them in a similar way. Under this condition, the SHAP values, while still reflecting the contribution of each variable to the model prediction, does not inform us on the minimum set of variables we are interested in in order to gain intuition on physics behind the model. 

\section{Constraining astrophysical parameters}
\label{sec:astro_params}

\begin{figure*}
\centering
\includegraphics[width=0.99\linewidth]{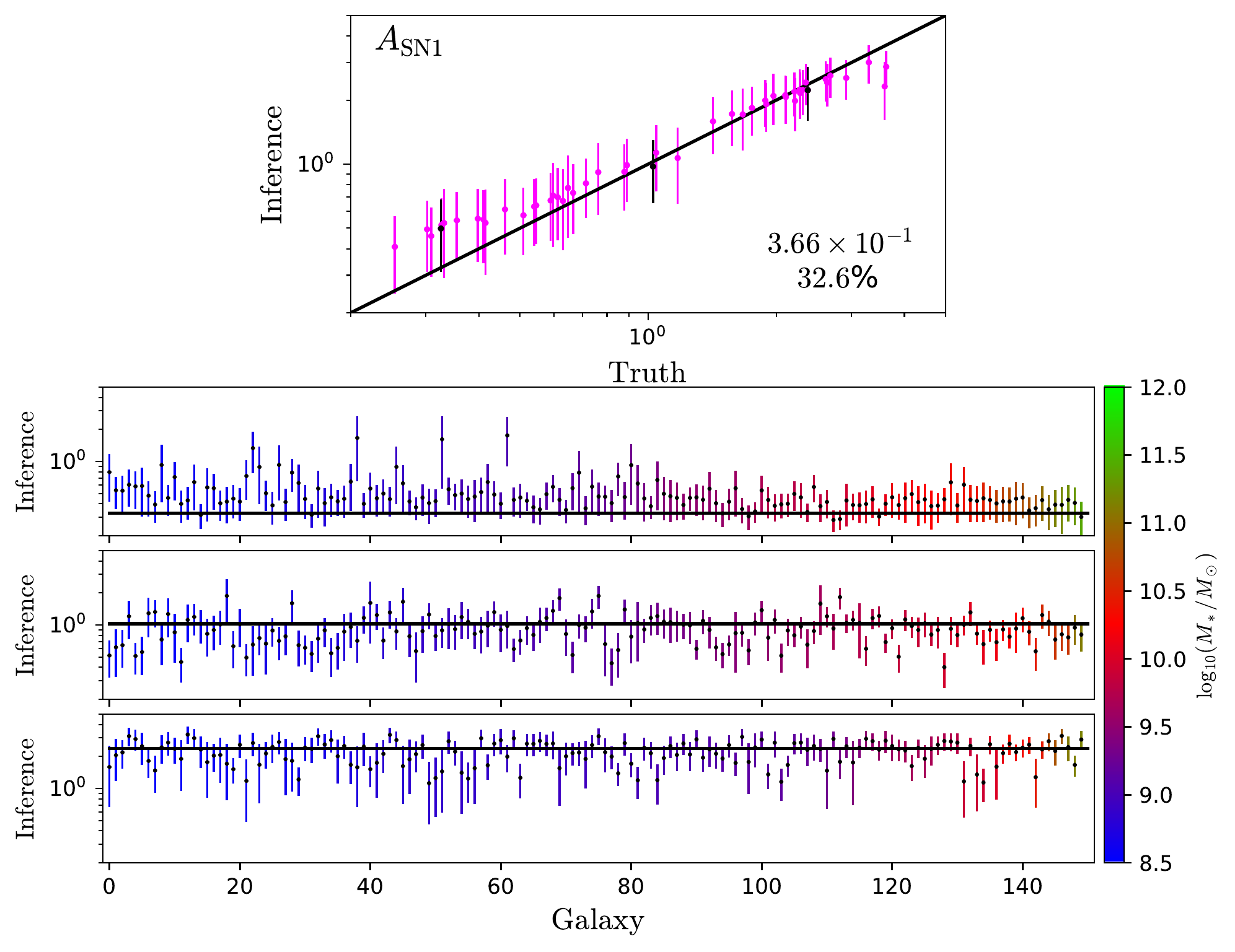}
\caption{Same as Fig. \ref{fig:IllustrisTNG_big} but predicting the astrophysical parameter $A_{\rm SN1}$ from IllustrisTNG.}
\label{fig:big_ASN1}
\end{figure*}

\begin{figure*}
\centering
\includegraphics[width=0.99\linewidth]{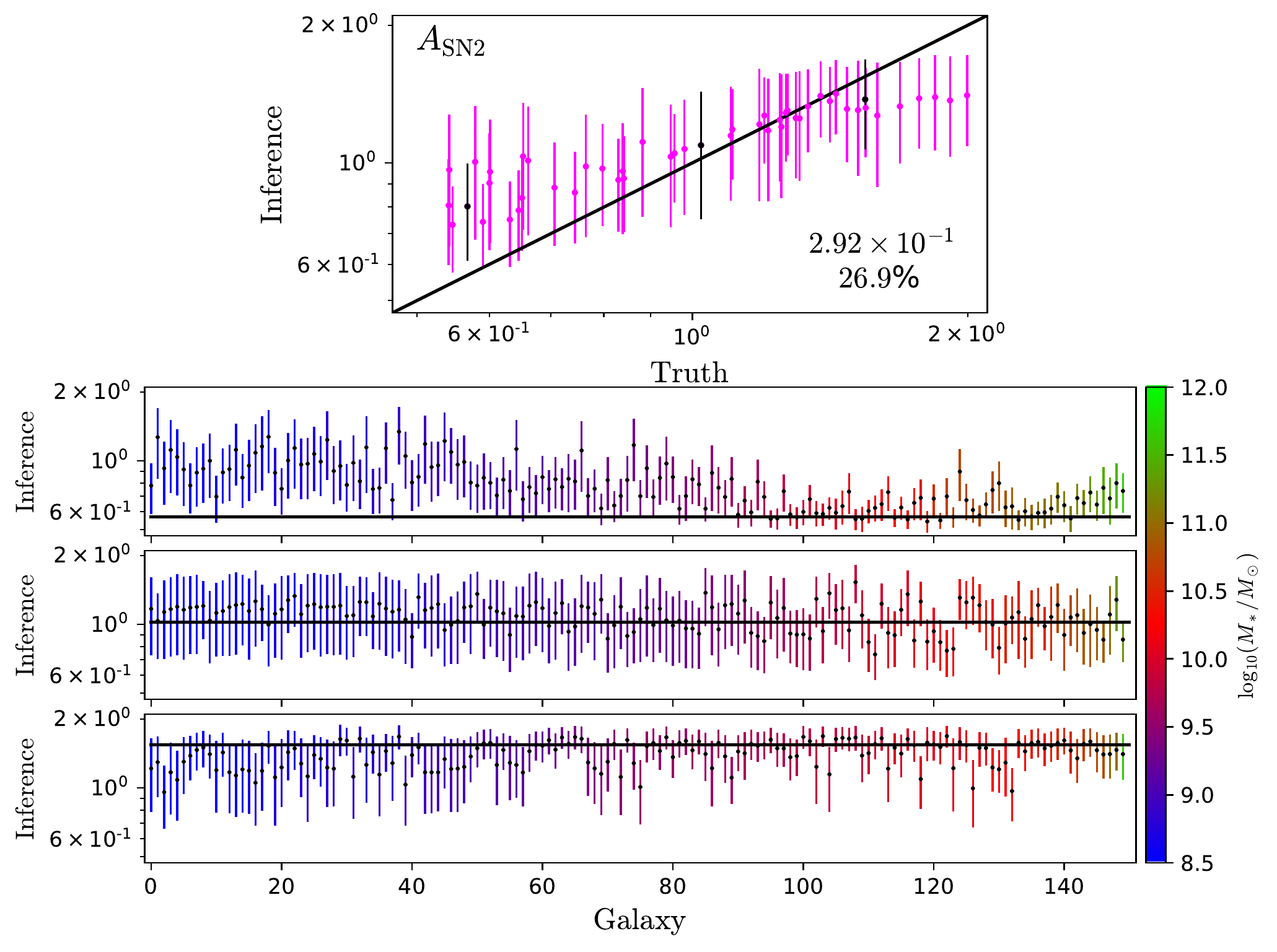}
\caption{Same as Fig. \ref{fig:IllustrisTNG_big} but predicting the astrophysical parameter $A_{\rm SN2}$ from IllustrisTNG.}
\label{fig:big_ASN2}
\end{figure*}

In this paper we have focused our attention in predicting the value of $\Omega_{\rm m}$. However, we saw in Fig. \ref{fig:results_IllustrisTNG} that our models seem to be able to have some constraining power on $A_{\rm SN1}$ and, to a lesser extend, on $A_{\rm SN2}$. In order to investigate this more, we have repeated the analysis outlined in Sec. \ref{sec:results} using IllustrisTNG galaxies at $z=0$ and show the results in Figs. \ref{fig:big_ASN1} and \ref{fig:big_ASN2} for the parameters $A_{\rm SN1}$ and $A_{\rm SN2}$, respectively.

For $A_{\rm SN1}$ we find that the model is able to infer its value with an accuracy of $\sim0.37$ and a precision of $\sim33\%$. On the other hand, for $A_{\rm SN2}$ the model can constrain its value with an accuracy and precision of $\sim0.29$ and $\sim27\%$, respectively. We note that although the numbers are better for $A_{\rm SN2}$, the visual inspection of the results reveals that these are largely affected by priors and the model actually performs better on $A_{\rm SN1}$.

When inspecting the results from individual galaxies more closely in the bottom panels of Figs. \ref{fig:big_ASN1} and \ref{fig:big_ASN2} we find that the model performs relatively well for $A_{\rm SN1}$ in general, while for $A_{\rm SN2}$ we can see that in many cases the model is just predicting the mean value with large errorbars, independently of galaxy type, cosmology, and astrophysics. 

From this exercise we conclude that while the network is capable of using galaxy properties to infer the value of $A_{\rm SN1}$ with large errorbars, it barely can say anything beyond predicting the mean value for $A_{\rm SN2}$. We emphasize that the network cannot infer the value of the other parameters not mentioned in this appendix, i.e. $A_{\rm AGN1}$, $A_{\rm AGN2}$, and $\sigma_8$.

\bibliography{references}{}
\bibliographystyle{aasjournal}

\end{document}